\documentclass[a4paper, 10pt, twoside, reqno]{article}
\usepackage{amsthm, amsfonts, xcolor, MnSymbol, hyperref, rotating, colortbl, tabularx, appendix, bm}

\usepackage{listings}
\definecolor{codegreen}{rgb}{0,0.6,0}
\definecolor{codegray}{rgb}{0.5,0.5,0.5}
\definecolor{codepurple}{rgb}{0.58,0,0.82}
\definecolor{backcolour}{rgb}{0.95,0.95,0.92}
\lstdefinestyle{mystyle}{
	backgroundcolor=\color{backcolour}, 
	commentstyle=\color{codegreen},
	keywordstyle=\color{magenta},
	numberstyle=\tiny\color{codegray},
	stringstyle=\color{codepurple},
	basicstyle=\ttfamily\tiny,
	breakatwhitespace=false,   
	breaklines=true,     
	captionpos=b,     
	keepspaces=true,     
	numbers=left,     
	numbersep=5pt,     
	showspaces=false,    
	showstringspaces=false,
	showtabs=false,     
	tabsize=2
}
\usepackage{geometry, enumerate}
\numberwithin{equation}{section} 
\newtheorem{theorem}{Theorem}[section]
\newtheorem{example}[theorem]{Example}
\newtheorem{corollary}[theorem]{Corollary}
\newtheorem{definition}[theorem]{Definition}
\newtheorem{proposition}[theorem]{Proposition}
\newtheorem{lemma}[theorem]{Lemma}
\newtheorem{remark}[theorem]{Remark}
\begin{document}
	\begin{center}
		\LARGE {\textbf{MDS multi-twisted Reed-Solomon codes with small dimensional hull}}
	\end{center}
	\begin{center}
		{Harshdeep Singh and Kapish Chand Meena\footnote{Corresponding Author E-mail: meenakapishchand@gmail.com}}
	\end{center}

\begin{abstract}	
	In this paper, we find a necessary and sufficient condition for multi-twisted Reed-Solomon codes to be MDS. 
	In particular, we introduce a new class of MDS double-twisted Reed-Solomon codes $\mathcal{C}_{\bm \alpha, \bm t, \bm h, \bm \eta}$ with twists $\bm t = (1, 2)$ and hooks $\bm h = (0, 1)$ over the finite field $\mathbb{F}_q$, providing a non-trivial example over $\mathbb{F}_{16}$ and enumeration over the finite fields of size up to 17.  
	Moreover, we obtain necessary conditions for the existence of multi-twisted Reed-Solomon codes with small dimensional hull. Consequently, we derive conditions for the existence of MDS multi-twisted Reed-Solomon codes with small dimensional hull.\\ 
	
\end{abstract}


\section{Introduction}
A linear $[n,k,d]$ code is said to be \textit{maximum distance separable} (MDS) code if its parameters achieve the Singleton bound \cite{MacWilliams1977}, i.e. $d=n-k+1$.
\textit{Reed-Solomon} (RS) and \textit{generalized Reed-Solomon} (GRS) codes \cite{Reed1960} are famous examples of MDS codes. For other families of MDS codes, we refer to \cite{MacWilliams1977}, \cite{Roth1989} and \cite{Roth1985}.
In \cite{Beelen2017}, Beelen at el. introduced MDS single-twisted RS codes as a generalization of RS codes.
The idea was to choose special type polynomials of degree $k$ with twist $t =1$ and hook $h =0, k-1$ such that those polynomials could have at most $k-1$ roots among the evaluation points.
Further, in \cite{Beelen2018a}, Beelen at al. extended the notion of single-twisted RS codes to multi-twisted RS codes by adding extra monomials (or twists) in the polynomials.
In recent years, applications of multi-twisted RS codes in code-based cryptography are studied by many authors (see: \cite{Beelen2018a}, \cite{Lavauzelle2020}).	
Hence, it is significant to explore algebraic criteria of multi-twisted RS codes to be MDS.

We know that dual $\mathcal{C}^{\bot}$ of a linear code $\mathcal{C}$ is also a linear code over the same finite field.
The hull of $\mathcal{C}$ is denoted by $Hull(\mathcal{C})$ and defined as $Hull(\mathcal{C})= \mathcal{C} \cap \mathcal{C}^{\bot}$.
It is clear that $Hull(\mathcal{C})$ is also a linear code.
There are many applications of the hull, e.g. determining the complexity of algorithms for checking equivalence of linear codes, computing the automorphism group of a linear code. It is also useful in side-channel attacks and fault injection attacks (see: \cite{Carlet2015, Leon1982, Leon1991, Luo2019, Sendrier1997, Sendrier2000, Sendriera}).
In \cite{Massey1992}, Massey introduced \textit{linear codes with complementary dual} (LCD), i.e. the codes having zero-dimensional hull. 
For other constructions of LCD codes we refer to \cite{Yang1994, Beelen2018, Carlet2018a}.
Moreover, Liu and Liu studied MDS LCD single-twisted RS codes and LCD multi-twisted RS codes (under certain assumptions) in \cite{Liu2021}. 
Linear codes with one-dimensional hull were studied in \cite{Li2019}.
In \cite{Wu2021}, Wu studied single-twisted RS codes with one-dimensional hull.

Motivated by their work, in this paper, we study the multi-twisted RS codes with general twists and hooks having zero and one-dimensional hull. First, we give necessary and sufficient condition for multi-twisted RS code to be MDS. Then we introduce new $k$-dimensional MDS double-twisted RS codes $\mathcal{C}_{\bm \alpha, \bm t, \bm h, \bm \eta}$ by choosing suitable $\bm \eta$ in twisted polynomials of degree $k+1$ with twists $\bm t = (1, 2)$ and hooks $\bm h = (0, 1)$ such that those polynomials could have at most $k-1$ roots among the evaluation points $E_{\bm\alpha}$.
Moreover, we obtain necessary conditions for the existence of multi-twisted Reed-Solomon codes with small dimensional hull. 
Then as a consequence, we derive conditions for the existence of MDS multi-twisted RS codes with small dimensional hull.

This paper is organized as follows: in Section \ref{sec2}, we recall some basic terminologies and results which are needed for this paper.
In Section \ref{sec3}, we describe a necessary and sufficient condition for a multi-twisted RS code to be MDS.
In Section \ref{sec4}, we explicitly describe the algebraic properties of MDS double-twisted RS codes over the finite fields $\mathbb{F}_q$, providing non-trivial examples over $\mathbb{F}_{16}$ and enumeration over the finite fields of order up to 17.
In Section \ref{sec5}, we study multi-twisted RS codes with small dimensional hull and provide examples. Also, we obtain necessary and sufficient conditions for the existence of MDS multi-twisted RS codes with small dimensional hull. Finally, in Section \ref{sec6}, we conclude with some future directions and compare these results with the existing results. 
%%%%%%%%%%%%%%%%%%%%%%%%%%%%%%%%%%%%%%%%%%%%%%%%%%%%%%%%%%%%%%%%%%%%%
%%%%%%%%%%%%%%%%%%%%%%%%%%%%%%%%%%%%%%%%%%%%%%%%%%%%%%%%%%%%%%%%%%%%%
\section{Preliminaries}\label{sec2}
%%%%%%%%%%%%%%%%%%%%%%%%%%%%%%%%%%%%%%%%%%%%%%%%%%%%%%%%%%%%%%%%%%%%%
%%%%%%%%%%%%%%%%%%%%%%%%%%%%%%%%%%%%%%%%%%%%%%%%%%%%%%%%%%%%%%%%%%%%%
In this section, we recall some fundamental concepts from literature which are needed for this paper.
\begin{definition} \cite{MacWilliams1977}
	Let $\alpha_1, \alpha_2, \dots, \alpha_n \in \mathbb{F}_q$ be distinct elements and $0 \leq k<n$. Then RS code is defined as
	\begin{equation}\label{eqn1}
		\mathcal{C}_{n,k}^{RS} = \{( f(\alpha_1), f(\alpha_2), \dots, f(\alpha_n) )\},
	\end{equation}
	where $f(x) \in \mathbb{F}_q[x]$ with $\deg(f(x))<k$.
\end{definition}
\noindent Here $\alpha_1, \alpha_2, \dots, \alpha_n$ are called evaluation points for the code. 
Clearly, $\mathcal{C}_{n,k}^{RS}$ is an $\mathbb{F}_q$-linear subspace of $\mathbb{F}_q^n$ having dimension $k$, i.e. $\mathcal{C}_{n,k}^{RS}$ is a linear code. 
Since the polynomial $f(x)$ has at most $k-1$ roots in the field, RS codes achieve the equality in the Singleton bound.

In \cite{Beelen2017}, Beelen et al. introduced the notion of twisted polynomials and using such polynomials they introduced single-twisted RS code. Let $0 \leq h < k \leq q-1$ and $\eta$ be a non-zero element of a finite field $\mathbb{F}_q$. 
Then the set of $(k,t,h,\eta)$-twisted polynomials over $\mathbb{F}_q$, is defined as 
\begin{equation*}
	\mathcal{V}_{k,t,h,\eta}:= \left \{f = \sum\limits_{i=0}^{k-1}a_i x^i + \eta a_h x^{k-1+t}: a_i \in \mathbb{F}_q \text{ for each }i\right \},
\end{equation*}
where $k, t, h\in \mathbb{N}$.
\begin{definition} $\cite{Beelen2017}$
	Let $\alpha_1, \alpha_2, \dots, \alpha_n \in \mathbb{F}_q$ be distinct elements and $\bm{\alpha} = (\alpha_1, \alpha_2, \dots, \alpha_n)$. Let $k<n$ and $0 < t \leq n-k$. Then the single-twisted RS code is defined as
	\begin{equation}\label{eqn2}
		\mathcal{C}_k(\bm\alpha,t,h,\eta) = \{(f(\alpha_1), f(\alpha_2), \dots, f(\alpha_n) )\},
	\end{equation}
	where $f(x) \in \mathcal{V}_{k,t,h,\eta}$ with $\deg(f(x))\leq k-1+t <n$.
\end{definition}
\noindent Note that, when $\eta=0$, then single-twisted RS codes become RS codes.
Since $\deg(f(x)) \not<k$, single-twisted RS codes are not MDS in general. However, in \cite{Beelen2017}, Beelen et al. obtained a necessary and sufficient condition for single-twisted RS codes to be MDS. Using SageMath implementations, they concluded that many long MDS single-twisted RS codes can be obtained using twisted polynomials for particular values of $q, n, k, h$ and  $t$; but not for general values. In addition, they claimed that in order to find explicit constructions of such long MDS single-twisted RS codes is difficult; however, they gave following result in favour to the existence.
\begin{lemma}(\cite{Beelen2017}, Theorem 17) 
	Let $\mathbb{F}_s \subset \mathbb{F}_q$ be a proper subfield and $\alpha_1, \alpha_2, \dots, \alpha_n \in\mathbb{F}_s$. If $\eta \in \mathbb{F}_q \setminus\mathbb{F}_s$, then the single-twisted code $\mathcal{C}_k(\bm\alpha,t,h,\eta)$ is MDS.
\end{lemma}
Further, in \cite{Beelen2018a}, Beelen at el. introduced multi-twisted RS codes as a generalization of single-twisted RS codes. 
They extended the twists and hooks: let $0 \leq h_i< h_{i+1} < k \leq q$ and $1\leq t_i <t_{i+1} \leq n-k$ for each $i$. 
Then, for $\bm{t} = (t_1,t_2,\dots,t_{\ell})$, $\bm{h}=(h_1,h_2,\dots,h_{\ell})$ and $\bm\eta= (\eta_1,\eta_2,\dots, \eta_{\ell}) \in (\mathbb{F}_q^{\ast})^\ell$, set of $(k,\bm{t}, \bm{h}, \bm{\eta})$-twisted polynomials over $\mathbb{F}_q$, is defined as
\begin{equation*}
	\mathcal{V}_{k, \bm{t}, \bm{h}, \bm{\eta}}:= \left \{f = \sum\limits_{i=0}^{k-1}a_i x^i + \sum\limits_{j=1}^\ell \eta_j a_{h_j} x^{k-1+t_j}: a_i \in \mathbb{F}_q \text{ for each }i \right \}.
\end{equation*}

\begin{definition}\label{dfn2.3}
	\cite{Beelen2018a}
	Let $\alpha_1, \alpha_2, \dots, \alpha_n $ be distinct elements of a finite field $\mathbb{F}_q$ and $k<n$. Then multi-twisted RS code is defined as
	\begin{equation}\label{eqn3}
		\mathcal{C}_k(\bm\alpha,\bm{t}, \bm{h}, \bm{\eta}) := \{(f(\alpha_1), f(\alpha_2), \dots, f(\alpha_n) )\},
	\end{equation}
	where $f(x) \in \mathcal{V}_{k, \bm{t}, \bm{h}, \bm{\eta}}$ with $\deg(f(x))\leq k-1+ t_\ell <n$.
\end{definition}
\noindent As per the Definition \ref{dfn2.3}, the generator matrix of multi-twisted RS code $\mathcal{C}_k(\bm\alpha,\bm{t}, \bm{h}, \bm{\eta})$ is
\begin{equation}\label{genmat}
	\left(\begin{array}{cccc}
		1 & 1 & \cdots & 1\\
		\alpha_1 & \alpha_2 & \cdots & \alpha_n \\
		\vdots & \vdots & \cdots & \vdots \\
		\alpha_1^{h_1}+ \eta_1 (\alpha_1)^{k-1+t_1} & \alpha_2^{h_1}+ \eta_1 (\alpha_2)^{k-1+t_1} & \cdots & \alpha_n^{h_1}+ \eta_1 (\alpha_n)^{k-1+t_1} \\
		(\alpha_1)^{h_1+1}&(\alpha_2)^{h_1 +1}&\cdots&(\alpha_n)^{h_1 +1}\\
		\vdots &\vdots &\vdots &\vdots \\
		\alpha_1^{h_2}+ \eta_2 (\alpha_1)^{k-1+t_2} & \alpha_2^{h_2}+ \eta_2 (\alpha_2)^{k-1+t_2} & \cdots & \alpha_n^{h_2}+ \eta_2 (\alpha_n)^{k-1+t_2}\\
		(\alpha_1)^{h_2+1}&(\alpha_2)^{h_2 +1}&\cdots&(\alpha_n)^{h_2 +1}\\
		\vdots &\vdots &\vdots &\vdots \\
		\alpha_1^{h_{\ell}}+ \eta_{\ell} (\alpha_1)^{k-1+t_{\ell}} & \alpha_2^{h_{\ell}}+ \eta_{\ell} (\alpha_2)^{k-1+t_{\ell}} & \cdots & \alpha_n^{h_{\ell}}+ \eta_{\ell} (\alpha_n)^{k-1+t_{\ell}}\\
		(\alpha_1)^{h_{\ell}+1}&(\alpha_2)^{h_{\ell} +1}&\cdots&(\alpha_n)^{h_{\ell} +1}\\
		\vdots &\vdots &\vdots &\vdots \\
		( \alpha_1)^{k-1} & ( \alpha_2)^{k-1} & \cdots & ( \alpha_n)^{k-1} \\
	\end{array}\right)_{k \times n}\begin{array}{c}
		\textcolor{white}{0}\\
		\textcolor{white}{0}\\
		\leftarrow (h_1 +1)^{th}~ row \\			
		\textcolor{white}{\vdots} \\
		\textcolor{white}{0}\\
		\leftarrow (h_2 +1)^{th} ~row \\
		\textcolor{white}{\vdots}\\
		\textcolor{white}{0}\\
		\leftarrow (h_{\ell} +1)^{th}~ row \\
		\textcolor{white}{\vdots}\\
		\textcolor{white}{0}\\
	\end{array}
\end{equation}
%%%%%%%%%%%%%%%%%%%%%%%%%%%%%%%%%%%%%%%%%%%%%%%%%%%%%%%%%%%%
%%%%%%%%%%%%%%%%%%%%%%%%%%%%%%%%%%%%%%%%%%%%%%%%%%%%%%%%%%%%
\section{MDS multi-twisted RS codes}\label{sec3}
%%%%%%%%%%%%%%%%%%%%%%%%%%%%%%%%%%%%%%%%%%%%%%%%%%%%%%%%%%%%
%%%%%%%%%%%%%%%%%%%%%%%%%%%%%%%%%%%%%%%%%%%%%%%%%%%%%%%%%%%%
In this section, we obtain a necessary and sufficient condition for a multi-twisted RS code to be MDS.
In \cite{Beelen2017}, authors gave the criteria for a single-twisted ($\ell=1$) RS code $\mathcal{C}_k(\bm \alpha, t,h,\eta)$ to be MDS.
We have extended their study for the multi-twisted RS code ($\ell > 1$) and give a more generalized structure as follows:
\begin{theorem}\label{lemma4.1}
	Let $\alpha_1, \alpha_2, \dots, \alpha_n \in \mathbb{F}_q$ be distinct and for $k<n<q$ and $\bm t=(t_1,t_2,\dots,t_\ell)$, $\bm h=(h_1,h_2,\dots,h_\ell)$ and $\bm \eta=(\eta_1,\eta_2,\dots, \eta_\ell)$ be as defined in previous section.
	Then the multi-twisted RS code $\mathcal{C}_{k}(\bm\alpha, \bm t,\bm h,\bm \eta)$ is MDS if and only if 
	for each $\mathcal{I} \subset \{1,2,\dots ,n\}$ with cardinality $k$ correspondingly the polynomial $\prod\limits_{i\in \mathcal{I}}(x-\alpha_i)= \sum\limits_{i} \sigma_i x^i$ with $\sigma_i =0$ for $i<0$,	
	the matrix
	\begin{equation*}
		\text{diag }\Big( \underset{\underset{t_{\ell}^{th}}{\uparrow} }{\eta_{\ell}^{-1}}, 1,\dots, 1, \underset{\underset{t_{\ell-1}^{th}}{\uparrow} }{\eta_{\ell-1}^{-1}}, 1,\dots, 1, \underset{\underset{t_{1}^{th}}{\uparrow} }{\eta_{1}^{-1}}, 1,\dots, \underset{\underset{1^{st}}{\uparrow} }{1} \Big)
		\cdot A_{\mathcal{I}} +B_\mathcal{I}
	\end{equation*}
	is non-singular. 
	Here, $A_\mathcal{I}$ and $B_\mathcal{I}$ are lower and upper-triangular $t_\ell\times t_\ell$ matrices, respectively, given by: 
	\begin{equation*}
		A_\mathcal{I} = \left(\begin{array}{lllll}
			1& 0 &0 & \cdots & 0\\
			\sigma_{k-1}& 1 & 0 &\cdots &0\\
			\sigma_{k-2}& \sigma_{k-1} & 1 &\cdots &0\\
			~~\vdots&~\vdots&~\vdots&~~\ddots&~\vdots\\
			\sigma_{k - t_\ell +1} & \cdots &\cdots & \sigma_{k-1}&1\\
		\end{array}\right) \text{ and}
	\end{equation*}
	{\footnotesize	
	\begin{equation*}
		B_\mathcal{I} = \left(\begin{array}{ccccccccc}
			-\sigma_{h_\ell-t_\ell+1}& -\sigma_{h_\ell-t_\ell+2} &\cdots & \cdots & \cdots & \cdots&\cdots&\cdots&\cdots\\
			0&\cdots&\cdots&\cdots&\cdots&\cdots&\cdots&\cdots&0\\
			
			\vdots&\vdots&\vdots&\vdots&\vdots&\vdots&\vdots&\vdots&\vdots\\
			
			0&\cdots&\cdots&\cdots&\cdots&\cdots&\cdots&\cdots&0\\
			0& \cdots&0&-\sigma_{h_{\ell-1}-t_{\ell-1}+1} & -\sigma_{h_{\ell-1}-t_{\ell-1}+2} &\cdots&\cdots&\cdots&\cdots \\
			
			0&\cdots&\cdots&0&\cdots&\cdots&\cdots&\cdots&0\\
			
			\vdots&\vdots&\vdots&\vdots&\vdots&\vdots&\vdots&\vdots&\vdots\\
			
			0&\cdots&\cdots&\cdots&\cdots&\cdots&\cdots&\cdots&0\\
			0& \cdots&\cdots&\cdots& 0 &-\sigma_{h_{1}-t_{1}+1} & -\sigma_{h_{1}-t_{1}+2} &\cdots&\cdots \\
			0&\cdots&\cdots&\cdots&\cdots&0&\cdots&\cdots&0\\
			
			\vdots&\vdots&\vdots&\vdots&\vdots&\vdots&\vdots&\vdots&\vdots\\
			
			0&\cdots&\cdots&0&\cdots&0&\cdots&\cdots&0\\
		\end{array}\right)
		\begin{array}{l}
			\leftarrow t_\ell^{th}\\
			\\
			\\
			\\
			\leftarrow t_{\ell-1}^{th}
			\\
			\\
			~~~\vdots\\
			\\
			\leftarrow t_1^{th}\\
			\\
			\\
			\leftarrow 1^{st}\\
		\end{array}
	\end{equation*}
	\begin{equation*}
		\begin{array}{ccccccccc}
			\uparrow~~~~~~~~~~~&&&~~~~~~~~~~~\uparrow~~~~~~~~~~~~~&&~~~~~~~~~~~\uparrow~~~~~~~~~~&&&~~~~~~~\uparrow\\
			t_\ell^{th}~~~~~~~~~~~~&&&~~~~~~~~~~~t_{\ell-1}^{th}~~~~~~~~~~~~~&\cdots&~~~~~~~~~~~t_1^{th}~~~~~~~~~~&&&~~~~~~~1^{st}\\
		\end{array}
	\end{equation*}
}
\end{theorem}
\begin{proof}
	It suffices to show that the codeword with at least $k$ zero positions is precisely the zero codeword. Let $f\in \mathcal{V}_{k, \bm{t}, \bm{h}, \bm{\eta}}$ corresponds to the codeword with at least $k$ zero positions, i.e. let $f$ have $k$ roots among the evaluation points. 
	Then we can write 
	\begin{equation}\label{fgsigma}
		f(x)= \sigma (x) g(x),
	\end{equation}
	where,
	$\sigma (x)=\prod\limits_{i \in \mathcal{I}} (x-\alpha_i) = \sum\limits_{i=0}^k \sigma_i x^i$ for some $\mathcal{I} \subset \{1,2,\dots,n\}$ with cardinality $k$ and $g(x) = \sum\limits_{i=0}^{t_\ell-1} g_i x^i\in \mathbb{F}_q[x]$. 
	Note that $\sigma_k=1$ and all the coefficients of $x^k$ to $x^{k+t_\ell-1}$ are zero except the coefficients of $x^{k+t_1-1}$, $x^{k+t_2-1}$, $\dots$, $x^{k+t_{\ell-1}-1}$, $x^{k+t_{\ell}-1}$ in $f$. 
	Consequently, we obtain the following system of $(t_\ell - \ell)$ equations in $g_j$s:
	\begin{equation}\label{eqn9}
		\sum\limits_{j=0}^i \sigma_{i-j}g_j =0,
	\end{equation}
	where $i \in \{k, k+1, \dots, k+t_{\ell}-2\} \setminus\{ k+t_1-1, k+t_2-1, \dots, k+t_{\ell-1}-1\}$ and $g_j=0$ when $j \notin \{0, 1, \dots, t_\ell -1\}$ and $\sigma_j=0$ when $j \notin \{0, 1, \dots, k\}$.
	Comparing the coefficients of $x^{k+t_s-1}$ in both sides of (\ref{fgsigma})
	and substituting the value of $a_{h_s} = \sum\limits_{j=0}^{h_s} \sigma_{h_s-j}g_j$, we obtain the following system of $\ell$ equations in $g_j$s:
	\begin{equation}\label{eqn10}
		\eta_{s}^{-1} \left(\sum\limits_{i=0}^{k}
		\sigma_{k-i}g_{t_s -1+i} \right) - \sum\limits_{j=0}^{h_s} \sigma_{h_s-j}g_j =0,
	\end{equation}
	for $s=1, 2, \dots, \ell$. 
	From (\ref{eqn9}) and (\ref{eqn10}), we have a homogeneous system of $t_\ell$ equations in $t_\ell$ variables. 
	The code $\mathcal{C}_{k}(\bm\alpha, \bm t,\bm h,\bm \eta)$ is MDS if and only if $f$ is zero polynomial, i.e. the above homogeneous system has only the zero vector as solution for all choices of $\mathcal{I}$. This completes the proof.
\end{proof}

\begin{example}\label{ex::3.2}
		Consider the finite field $\mathbb{F}_{2^4}= \mathbb{F}_2(\alpha)$ with $\alpha^4 + \alpha +1= 0$.
		Let $n=5$, $k=3$, $\bm \alpha= (0, \alpha^2, \alpha + 1, \alpha^2 + \alpha, \alpha^3 + \alpha + 1)$, $\bm t= (1,2)$ and $\bm h = (0,1)$. Let $\mathcal{C}_{k}(\bm\alpha, \bm t,\bm h,\bm \eta)$ be a multi-twisted RS code having $\ell=2$ twists with $\bm \eta = (\eta_1,\eta_2)$, where $\eta_1= \alpha^3 + \alpha^2$ and $\eta_2 \in \{1, \alpha^2 + 1, \alpha^2 + \alpha + 1, \alpha^3, \alpha^3 + \alpha^2, \alpha^3 + \alpha^2 + \alpha\}$. Now for each $\mathcal{I}\subset\{1,2,3,4,5\}$ with cardinality $3$ correspondingly polynomial $\prod\limits_{i\in\mathcal{I}}(x-\alpha_i)= \sum\limits_i\sigma_ix^i$, the matrix 
		\begin{equation*}
			\begin{bmatrix}
				\eta_2^{-1}&0\\
				0& \eta_1^{-1}
			\end{bmatrix}\cdot
			\begin{bmatrix}
				1 & 0\\ \sigma_2 & 1
			\end{bmatrix}+
		\begin{bmatrix}
			-\sigma_0& -\sigma_1\\ 0 & -\sigma_0
		\end{bmatrix}
		\end{equation*}
		is non-singular. Then by Theorem \ref{lemma4.1} the code $\mathcal{C}_{k}(\bm\alpha, \bm t,\bm h,\bm \eta)$ is MDS. For detailed implementation we refer to Appendix \ref{Appendix1}.
\end{example}

\begin{remark}
	By SageMath implementations, similar to the existence of MDS single-twisted RS codes one can see that many long MDS multi-twisted RS codes can be obtained using twisted polynomials for particular values of $q, n, k, \bm h$ and  $\bm t$; but not for general values. Also, it seems hard to find explicit constructions of such long MDS multi-twisted RS codes. However, the existence of MDS multi-twisted RS code given by Beelen et al. \cite{Beelen2018a}, uses the fact that $\mathcal{C}_k(\bm\alpha, \bm t,\bm h,\bm \eta)$ is MDS if and only if every $k$ columns of generator matrix (\ref{genmat}) are linearly independent over $\mathbb{F}_q$.
	For this, the authors assumed some constraints on the evaluation vector $\bm\alpha$ and the vector $\bm \eta$. 
	Now, we give an alternate proof of the same result as a consequence of the Theorem \ref{lemma4.1}.
\end{remark}

\begin{theorem}\label{ncforexistenceofeta} 
	\cite{Beelen2018a}
	Let $\mathbb{F}_{q_0} \subsetneq \mathbb{F}_{q_1}
	\subsetneq \mathbb{F}_{q_2}
	\subsetneq \cdots
	\subsetneq \mathbb{F}_{q_\ell}
	=	
	\mathbb{F}_q$ be a proper chain of subfields and $\alpha_1, \alpha_2, \dots, \alpha_n \in \mathbb{F}_{q_0}$. 
	If $\eta_{i} \in \mathbb{F}_{q_i} \setminus \mathbb{F}_{q_{i-1}}$, for all $1 \leq i \leq \ell$. Then the multi-twisted RS code $\mathcal{C}_{k}(\bm\alpha, \bm t,\bm h,\bm \eta)$ is MDS.
\end{theorem}
\begin{proof}
	Let 
	$\mathcal{I} \subset \{1,2,\dots,n\}$ with cardinality $k$
	and $ W =\text{diag}\Big(\eta_{\ell}^{-1}, 1,\dots, 1, \eta_{\ell-1}^{-1}, 1,\dots, 1, \eta_{1}^{-1}, 1,\dots, 1 \Big)
	\cdot A_{\mathcal{I}} +B_\mathcal{I}$ be the corresponding matrix as described in Theorem \ref{lemma4.1}. 
	Since $\alpha_i \in \mathbb{F}_{q_0}$ for each $i$, $\sigma_j$ also belongs to $\mathbb{F}_{q_0}$ for each $j$.
	In particular, $\sigma_0 \in \mathbb{F}_{q_0}$. 
	Further, $\eta_j^{-1} \in \mathbb{F}_{q_j} \setminus \mathbb{F}_{q_{j-1}}$ implies that $\eta_j^{-1} - \sigma_j \neq 0$.
	Hence, the diagonal entries in $W$ are all non-zero and on applying elementary row operations, we can convert this matrix to lower triangular matrix with diagonal elements $\eta_\ell^{-1}+T_\ell, 1, \dots, 1, \eta_{\ell-1}^{-1}+T_{\ell-1}, 1, \dots,\dots ,1, \eta_{1}^{-1}+T_{1}, 1, \dots, 1$ for some $T_i \in \mathbb{F}_{q_{i-1}}$.
	Hence, corresponding matrix is non-singular. 
	Thus by Theorem \ref{lemma4.1}, $\mathcal{C}_{k}(\bm\alpha, \bm t,\bm h,\bm \eta)$ is MDS. This completes the proof.
\end{proof}
%%%%%%%%%%%%%%%%%%%%%%%%%%%%%%%%%%%%%%%%%%%%%%%%%%%%%%%%%%%

%%%%%%%%%%%%%%%%%%%%%%%%%%%%%%%%%%%%%%%%%%%%%%%%%%%%%%%%%%%
\section{MDS double-twisted RS codes}\label{sec4}
In this section, we give algebraic criteria for double-twisted RS codes having twists $\bm t = (1, 2)$ and hooks $\bm h = (0, 1)$ to be MDS with enumeration.
For these double-twisted RS codes, we assume $\bm \eta = (\eta_1, \eta_2) \in (\mathbb{F}_q^*)^2$. 
Let $k <n<q$ then we have $\mathcal{V}_{k,\bm t,\bm h,\bm \eta} :=\{a_0+ a_1 x + \cdots + a_{k-1}x^{k-1} + \eta_1 a_0 x^k + \eta_2 a_1 x^{k+1} : a_i \in \mathbb{F}_q \text{ for each }i\}$. Let $\bm{\alpha}= (\alpha_1, \alpha_2, \dots, \alpha_n) \in \mathbb{F}_q^n$ be an evaluation vector and $E_{\bm \alpha} = \{\alpha_1, \alpha_2, \dots, \alpha_n\}\subset \mathbb{F}_q$ be set of evaluation points, where $\alpha_i \neq \alpha_j$ for each $i \neq j$. 

\begin{proposition}\label{prop::4.1}
	The polynomial $f(x) = a_0+ a_1 x + \cdots + a_{k-1}x^{k-1} + \eta_1 a_0 x^k + \eta_2 a_1 x^{k+1} \in \mathcal{V}_{k,\bm t,\bm h,\bm \eta}$ has at most $k-1$ roots among $E_{\bm \alpha}$ if any of the following conditions hold:
	\begin{enumerate}[(i)]
		\item $a_0=a_1=0$;
		\item $a_0 \neq 0$, $a_1=0$, and $\eta_1 \neq \frac{(-1)^k}{\prod\limits_{i \in \mathcal{J}_k}\alpha_{i}}$;
		\item $a_0 \neq 0$, $a_1\neq0$,
		$\eta_2 \neq \frac{(-1)^k}{\prod\limits_{i\in \mathcal{J}_k}\alpha_{i}}$, and 
		$\eta_1 \neq 
		\frac{
			\Big(
			\sum\limits_{i \in \mathcal{J}_{k}} \prod\limits_{\underset{j\neq i}{j \in \mathcal{J}_{k}} } \alpha_j
			\Big)
			\left(
			\sum\limits_{i \in \mathcal{J}_k} \alpha_{i}
			\right)+ 
			\left(
			\frac{(-1)^k}{\eta_2} - \prod\limits_{i \in \mathcal{J}_k}\alpha_{i}
			\right)}
		{(-1)^k\left(\prod\limits_{i \in \mathcal{J}_k}\alpha_{i}\right)
			\left(
			\frac{(-1)^k}{\eta_2} - \prod\limits_{i \in \mathcal{J}_k}\alpha_{i}
			\right)}$;
		
		\item $a_0=0$, $a_1 \neq0$,
		and $\eta_2 \neq \frac{(-1)^{k-1}}{\left(\sum\limits_{j\in \mathcal{J}_{k-1}}\alpha_j\right)\left(\prod\limits_{j\in \mathcal{J}_{k-1}}\alpha_j\right)}$;
	\end{enumerate}
	\noindent where $\mathcal{J}_{s} \subset \{1,2,\dots,n\}$ having cardinality $s$ such that $\alpha_{j} \neq0$ for each $j\in \mathcal{J}_s$.
	
	\begin{proof}
		It is immediate to observe that if $(i)$ or $(ii)$ holds, then $f$ has at most $k-1$ roots among $E_{\bm \alpha}$.
		If $(iii)$ holds then since $a_0 \neq 0$, $f$ doesn't have $0$ among its roots.
		Now, suppose  roots of $f$ are $\alpha_{i_1}, \alpha_{i_2}, \dots, \alpha_{i_k}$ and $\beta$, where $\alpha_{i_j}\in E_{\bm \alpha}$ for $1\leq j\leq k$. 
		Then $f(x) = \eta_2 a_1 (x-\beta) \prod_{j=1}^k (x-\alpha_{i_j})$, where $\eta_2\neq0$.  
		Now the product of roots of $f(x)$ is 
		\begin{equation}\label{eq::prop4.2::1}
			\beta	\prod\limits_{j=1}^k\alpha_{i_j} =\frac{(-1)^{k+1}a_0}{\eta_2 a_1}
		\end{equation}
		and, the sum of product of $k$-roots of $f$, i.e. 
		\begin{equation*}
			\prod\limits_{j=1}^k \alpha_{i_j}+ \beta\left(\alpha_{i_1} \alpha_{i_2} \cdots \alpha_{i_{k-1}} + \alpha_{i_1} \cdots \alpha_{i_{k-2}}\alpha_{i_{k}} + \cdots + \alpha_{i_2} \alpha_{i_3} \cdots \alpha_{i_{k}}\right) = \frac{(-1)^k}{\eta_2}.
		\end{equation*}
		Now if $\alpha_{i_1} \alpha_{i_2} \cdots \alpha_{i_{k-1}} + \alpha_{i_1} \cdots \alpha_{i_{k-2}}\alpha_{i_{k}} + \cdots + \alpha_{i_2} \alpha_{i_3} \cdots \alpha_{i_{k}}=0$, then $\eta_2 =\frac{(-1)^k}{\prod_{j=1}^k\alpha_{i_j}}$, which contradicts the assumption $(iii)$. 
		Therefore
		\begin{equation}\label{eq::prop4.2::2}
			\beta = \frac{\left(\frac{(-1)^k}{\eta_2} - \prod\limits_{j=1}^k\alpha_{i_j}\right)}{\left(\alpha_{i_1} \alpha_{i_2} \cdots \alpha_{i_{k-1}} + \alpha_{i_1} \cdots \alpha_{i_{k-2}}\alpha_{i_{k}} + \cdots + \alpha_{i_2} \alpha_{i_3} \cdots \alpha_{i_{k}}\right)}.
		\end{equation}
		Also the sum of roots of $f$ is
		\begin{equation}\label{eq::prop4.2::3}
			\sum\limits_{j=1}^k \alpha_{i_j}+ \beta = \frac{-\eta_1 a_0}{\eta_2a_1}.
		\end{equation}
		Using (\ref{eq::prop4.2::1}) and (\ref{eq::prop4.2::3}), we obtain
		\begin{equation}\label{eq::prop4.2::4}
			\sum\limits_{j=1}^k \alpha_{i_j}+ \beta = (-1)^k \eta_1 \left( \beta\prod\limits_{j=1}^k\alpha_{i_j} \right).
		\end{equation}
		Using (\ref{eq::prop4.2::2}) in (\ref{eq::prop4.2::4}), we obtain
		\begin{equation}\label{eta_1_not_equal}
			\eta_1 = \frac{\left(\alpha_{i_1} \alpha_{i_2} \cdots \alpha_{i_{k-1}} + \alpha_{i_1} \cdots \alpha_{i_{k-2}}\alpha_{i_{k}} + \cdots + \alpha_{i_2} \alpha_{i_3} \cdots \alpha_{i_{k}}\right)\left(\sum\limits_{j=1}^k \alpha_{i_j}\right)+ {\left(\frac{(-1)^k}{\eta_2} - \prod\limits_{j=1}^k \alpha_{i_j}\right)}}{ (-1)^k \left(\prod\limits_{j=1}^k \alpha_{i_j}\right){\left(\frac{(-1)^k}{\eta_2} - \prod\limits_{j=1}^k\alpha_{i_j}\right)} }
		\end{equation}
		but this contradicts the assumption $(iii)$. Note that this contradiction is free from the choices of $\beta$. Observe that in (\ref{eta_1_not_equal}) denominator is non-zero by (\ref{eq::prop4.2::1}) and (\ref{eq::prop4.2::2}). Hence $f$ can have at most $k-1$ roots among $E_{\bm \alpha}$.
		Lastly, assume $(iv)$ holds then since $a_0=0$, one of the roots of $f$ is definitely 0. This gives a factor of $f$; denoted by $f/x$ as $a_1 + a_2x + \cdots + a_{k-1}x^{k-2} + \eta_2 a_1 x^k$.
		Therefore, we have following two cases:
		\begin{itemize}
		\item $0 \notin E_{\bm \alpha}$: 
		Let roots of $f$ be $0,\alpha_{i_2},\alpha_{i_3},\dots, \alpha_{i_{k+1}}$, 
		then $\sum_{j=2}^{k+1}\alpha_{i_j}=0$ since sum of roots of $f$ is zero. That means
		\begin{equation}\label{sumofrootsoffwhen0isnotinealpha}
			\alpha_{i_2} = -\sum_{j=3}^{k+1}\alpha_{i_j}.
		\end{equation}
		Assume $\alpha_{i_j} \in E_{\bm\alpha}$ for each $j$, then $f/x$ also has these $k$ roots in $E_{\bm\alpha}$. 
		Then the product of roots of $f/x$ is 
		\begin{equation}\label{productofrootsoff/xwhen0isnotinealpha}
			\frac{(-1)^k}{\eta_2}=\prod_{j=2}^{k+1}\alpha_{i_j}.
		\end{equation}
		Using (\ref{sumofrootsoffwhen0isnotinealpha}) in (\ref{productofrootsoff/xwhen0isnotinealpha}), we get $\eta_2 = \frac{(-1)^{k-1}}{\left(\sum \limits_{j=3}^{k+1}\alpha_{i_j}\right)\left(\prod\limits_{j=3}^{k+1}\alpha_{i_j}\right)}$. In general we obtain  $\eta_2 = \frac{(-1)^{k-1}}{\left(\sum\limits_{j\in \mathcal{J}_{k-1}}\alpha_j\right)\left(\prod\limits_{j\in \mathcal{J}_{k-1}}\alpha_j\right)}$ where $\mathcal{J}_{s} \subset \{1,2,\dots,n\}$ having cardinality $s$ such that $\alpha_{j} \neq0$ for each $j\in \mathcal{J}_s$. This contradicts the assumption.
		\item $0 \in E_{\bm \alpha}$: Let roots of $f/x$ be $\alpha_{i_1},\alpha_{i_2},\dots,\alpha_{i_{k-1}},\beta$ where $\alpha_{i_j}\in E_{\bm\alpha}$. 
		The product and sum of roots of $f/x$ are given by
		\begin{align*}
			\beta \prod_{j=1}^{k-1} \alpha_{i_j} = \frac{(-1)^k}{\eta_2}~\text{and}~\beta + \sum_{j=1}^{k-1} \alpha_{i_j} = 0,~\text{respectively}.
		\end{align*}
		\noindent Now eliminating $\beta$ we obtain a contradiction to the condition in $(iv)$. Again note that this contradiction is free from the choices of $\beta$. Hence $f/x$ can have at most $k-2$ roots among $E_{\bm \alpha}$.
	\end{itemize}
	This completes the proof.	
		\end{proof}
\end{proposition}

\begin{theorem}\label{thm::double_twisted_rs}
	Let $k<n$ and $\bm \alpha = (\alpha_1, \alpha_2,\dots,\alpha_n)\in \mathbb{F}_q^n$ where $\alpha_i \neq \alpha_j$ for each $i\neq j$, $\bm t = (1,2)$, $\bm h = (0,1)$ and $\bm \eta = (\eta_1,\eta_2)\in {(\mathbb{F}_q^*)^{2}}$. Then the double-twisted RS code 
	\begin{equation*}
		\mathcal{C}_k(\bm \alpha, \bm t, \bm h, \bm \eta)= \{(f(\alpha_1), f(\alpha_2), \dots, f(\alpha_n))\in \mathbb{F}_q^n: f\in \mathcal{V}_{k,\bm t,\bm h,\bm \eta}\}
	\end{equation*}
	is MDS if and only if the following conditions hold:
	\begin{enumerate}[(i)]
		\item $\eta_1 \neq
		\frac{(-1)^k}{\prod\limits_{i \in \mathcal{J}_k}\alpha_{i}}$ whenever $\sum\limits_{i \in \mathcal{J}_{k}} \prod\limits_{\underset{j\neq i}{j \in \mathcal{J}_{k}} } \alpha_j = 0$;
		
		\item $\eta_1 \neq 
		\frac{
			\Big(
			\sum\limits_{i \in \mathcal{J}_{k}} \prod\limits_{\underset{j\neq i}{j \in \mathcal{J}_{k}} } \alpha_j
			\Big)
			\left(
			\sum\limits_{i \in \mathcal{J}_k} \alpha_{i}
			\right)+ 
			\left(
			\frac{(-1)^k}{\eta_2} - \prod\limits_{i \in \mathcal{J}_k}\alpha_{i}
			\right)}
		{(-1)^k\left(\prod\limits_{i \in \mathcal{J}_k}\alpha_{i}\right)
			\left(
			\frac{(-1)^k}{\eta_2} - \prod\limits_{i \in \mathcal{J}_k}\alpha_{i}
			\right)}$ whenever $\eta_2 \neq \frac{(-1)^k}{\prod\limits_{j\in \mathcal{J}_k}\alpha_j}$;
		
		\item $\eta_2 \neq \frac{(-1)^{k-1}}{\left(\sum\limits_{j\in \mathcal{J}_{k-1}}\alpha_j\right)\left(\prod\limits_{j\in \mathcal{J}_{k-1}}\alpha_j\right)}$ whenever $\prod\limits_{i\in \mathcal{I}_{k}}\alpha_i =0$
	\end{enumerate}
	\noindent where 
	$\mathcal{I}_s$ and 
	$\mathcal{J}_{s} \subset \{1,2,\dots,n\}$ having cardinality $s$ such that $\alpha_{j} \neq0$ for each $j\in \mathcal{J}_s$.

	\begin{proof}
		The code $\mathcal{C}_k(\bm \alpha, \bm t, \bm h, \bm \eta)$ is MDS if and only if every polynomial $f \in \mathcal{V}_{k,\bm t,\bm h,\bm \eta}$ has at most $k-1$ roots among the evaluation points $E_{\bm \alpha}$. 
		The proof of if part follows by Proposition \ref{prop::4.1}. On the other hand, if any of the conditions $(i)$-$(iii)$ do not hold then by similar approach to \cite{Beelen2017} one can find a polynomial $f \in \mathcal{V}_{k,\bm t,\bm h,\bm \eta}$ having at least $k$ roots among the evaluation points $E_{\bm \alpha}$. Thus we get required proof.
	\end{proof}
\end{theorem}

\begin{example}\label{ex::DTRS}
	Consider a finite field $ \mathbb{F}_{2^4}= \frac{\mathbb{F}_2[x]}{\left<x^4 + x +1\right>} = \mathbb{F}_2(\alpha)$. 
	Let $n = 5$, $k = 3$, $\bm t=(1,2)$, $\bm h=(0,1)$, and $\bm \eta = (\eta_1, \eta_2)\in \mathbb{F}_{2}(\alpha)^2$, then we have the set of double-twisted polynomials $\mathcal{V}_{k,\bm t,\bm h,\bm \eta}= \{a_0+ a_1 x + a_2 x^2 + \eta_1 a_0 x^3 + \eta_2 a_1 x^{4} : a_i \in \mathbb{F}_{2}(\alpha) \text{ for each }i\}$. 
	Then every $f \in \mathcal{V}_{k,\bm t,\bm h,\bm \eta}$ has at most $2$ roots among $\{0, \alpha^3 + \alpha^2, \alpha^3 + \alpha^2 + \alpha + 1, \alpha^3 + 1, 1\} \subset \mathbb{F}_{2}(\alpha)$ for $\eta_1 =\alpha^2 + \alpha$ and $\eta_2 \in \{1, \alpha, \alpha^2 + \alpha, \alpha^3, \alpha^3 + \alpha, \alpha^3 + \alpha^2\}$. 
	Thus the double-twisted RS code $\mathcal{C}_k(\bm \alpha,\bm t,\bm h,\bm \eta)$ is MDS.
\end{example}

\begin{remark}\label{remark4.4}
	If we apply the similar approach as Theorem 3.3 of \cite{Sui_2022, Sui_2022a} and Theorem 3.2 of \cite{Gu}, we obtain that the double-twisted RS code $\mathcal{C}_k(\bm \alpha, \bm t, \bm h, \bm \eta)$ in Theorem \ref{thm::double_twisted_rs} is MDS if and only if $\bm \eta = (\eta_1, \eta_2) \in (\mathbb{F}_q^*)^2$ satisfies
	\begin{align*}
		1 - \eta_1 (-1)^k \left( \prod\limits_{i \in \mathcal{I}_{k}} \alpha_i \right)+ \eta_2(-1)^k \left( \left(\sum_{j \in \mathcal{I}_k}\prod\limits_{\underset{i\in \mathcal{I}_k}{i \neq j}}\alpha_i\right)\left(\sum\limits_{i\in \mathcal{I}_{k}}\alpha_i\right) - \left(\prod\limits_{j \in \mathcal{I}_k}\alpha_j \right)\right) + \eta_1 \eta_2 \left(\prod\limits_{i \in\mathcal{I}_k} \alpha_i^2 \right) \neq 0
	\end{align*}
	for every $\mathcal{I}_{k} \subseteq \{1,2,\dots,n\}$ of cardinality $k$. One can easily verify that $\eta_1$ and $\eta_2$ obtained by Theorem \ref{thm::double_twisted_rs} satisfy the above condition and vice-versa. However, Theorem \ref{thm::double_twisted_rs} gives an approach to discard some choices for $\eta_1$ and $\eta_2$, which is better than exhaustive search.
\end{remark}
	Note that the MDS codes described in Theorem \ref{thm::double_twisted_rs} are double-twisted RS codes with twists $\bm t = (1, 2)$ and hooks $\bm h = (0, 1)$, which are particular multi-twisted RS codes with $\ell = 2$, and different from the MDS double-twisted RS codes obtained in \cite{Sui_2022a} and \cite{Gu}. Now we give the enumeration of such classes of MDS double-twisted RS codes over finite fields of size up to 17 in Table \ref{table::countingSTMDSRS}. By SageMath implementations, one can see that many long MDS double-twisted RS codes can be obtained using twisted polynomials for particular values of $q, n, k, \bm h$ and  $\bm t$; but not for general values. Also, it seems hard to find explicit constructions of such long MDS double-twisted RS codes. However, if we include additional conditions defined in Theorem \ref{ncforexistenceofeta} then MDS double-twisted RS codes exist always.

\begin{table}
	\begin{center}
		\includegraphics[width=16.5cm]{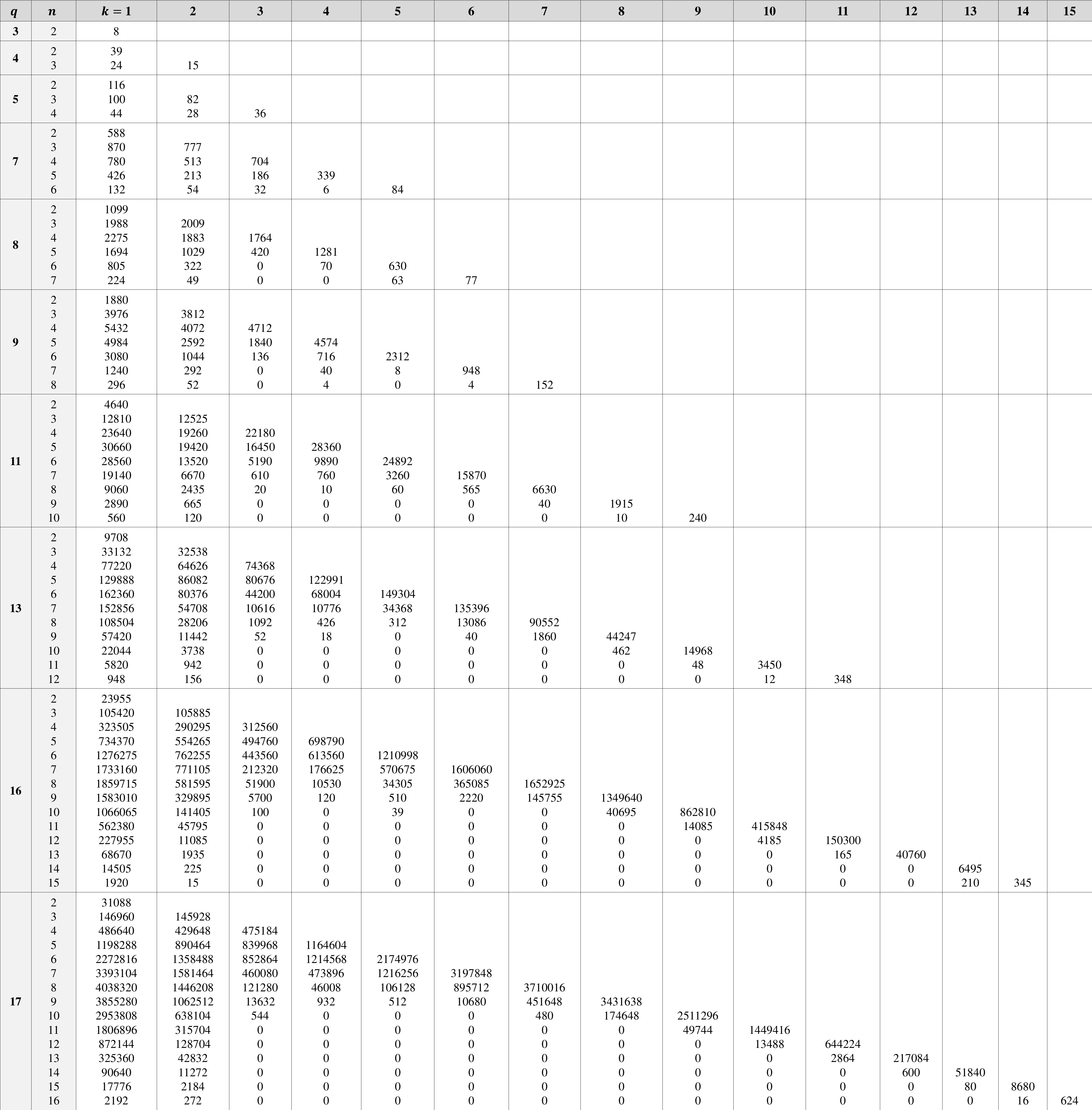}
	\end{center}
	\caption{\small Number of MDS double-twisted RS codes with twists $\bm t = (1,2)$ and hooks $\bm h = (0,1)$ respectively over finite fields of size $\leq 17$}\label{table::countingSTMDSRS}
\end{table}

\section{Multi-twisted RS codes with small dimensional hull}\label{sec5}

In this section, we study the existence of multi-twisted RS codes with small-dimensional hull. For this, we first state some useful results based on hull.
\begin{lemma}\label{lem4.1}
	($\cite{Li2019}$, Proposition 1) Let $\mathcal{C}$ be an $[n,k]$ linear code over $\mathbb{F}_q$ with generator matrix $G$. Then code $\mathcal{C}$ has a one-dimensional hull if and only if the rank of matrix $G \cdot G^T$ is $k-1$.
	Alternately, the code $\mathcal{C}$ has a one-dimensional hull if and only if the rank of matrix $H \cdot H^T$ is $n-k-1$.
\end{lemma}

\begin{proposition}\cite{Li2019}
	Let $\mathcal{C}$ be an $[n,k]$ linear code over $\mathbb{F}_q$ with generator matrix $G$. Then code $\mathcal{C}$ has a $\mathfrak{L}$-dimensional hull if and only if the rank of matrix $G \cdot G^T$ is $k-\mathfrak{L}$, where $0 
	\leq \mathfrak{L} \leq min\{k, n-k\}$.
\end{proposition}

\noindent In the remaining part of this section, we fix the following notations:\\ \medskip
\begin{tabular}{ll}
	$\gamma$& : a primitive element of $\mathbb{F}_q$\\
	$k$& : a positive integer such that $k\vert (q-1)$\\
	$\alpha_i$ &$ :=\gamma^{\frac{q-1}{k}i}$ for $1\leq i\leq k$\\
\end{tabular}\\
It is clear that $\gamma^{\frac{q-1}{k}}$ generates a subgroup of $\mathbb{F}_q^\ast$ of order $k$. 
Also, by \cite{Wu2021}, for any integer $m$, 
\begin{equation}\label{eq8}
	\theta^m=\alpha_1^m + \alpha_2^m + \cdots + \alpha_k^m = \left\{ \begin{array}{ll}
		k & \text{if }m \equiv 0 \bmod k\\
		0 & \text{ otherwise. }
	\end{array} \right.
\end{equation}
Next, we give proof for the existence of multi-twisted RS codes as defined in Definition \ref{dfn2.3}, with small dimensional hull in two scenarios when $q$ is even and odd.
\begin{theorem}\label{prop4.2}
	Let $q$ be a power of $2$, and
	$k>1$ be an integer such that $k \vert (q - 1)$.
	Then there exists a $[2k,k]$ multi-twisted RS code $\mathcal{C}_k(\bm\alpha,\bm{t},\bm{h},\bm{\eta})$ over $\mathbb{F}_q$ with $\mathfrak{L}$-dimensional hull ($\mathfrak{L}>0$) for $\bm\alpha = (\alpha_1,\dots,\alpha_k,\gamma\alpha_1,\dots,\gamma\alpha_k)$, $h_1>0$ and $t_1>1$.
	\begin{proof}
		Let, for $\beta \in \mathbb{F}_q^*$, $A_{\beta}$ be a $k \times k$ matrix over $\mathbb{F}_q$ given by 
		$$A_{\beta}=\left(\begin{array}{cccc}
			1 & 1 & \cdots & 1\\
			\beta \alpha_1 & \beta \alpha_2 & \cdots & \beta \alpha_k \\
			(\beta \alpha_1)^2 & (\beta \alpha_2)^2 & \cdots & (\beta \alpha_k)^2 \\
			\vdots &\vdots &\vdots &\vdots \\
			(\beta \alpha_1)^{k-1} & (\beta \alpha_2)^{k-1} & \cdots & (\beta \alpha_k)^{k-1} \\
		\end{array}\right),$$
		then, by using (\ref{eq8}), we have
		\begin{equation}\label{aat}
			A_{\beta}A_{\beta}^T=\left(\begin{array}{ccccc}
				k & 0 & \cdots & 0& 0\\
				0 & \cdots & \cdots &0& k \beta^k \\
				0 & \cdots & \cdots & k \beta^k&0 \\
				\vdots &\vdots &\dots &\vdots &\vdots \\
				0 & k \beta^k & 0 & \cdots & 0 \\
			\end{array}\right).
		\end{equation}
		Considering $B_{\beta}$ a $k \times k$ matrix over $\mathbb{F}_q$ as
		$$B_{\beta}=\left(\begin{array}{cccc}
			0 & 0 & \cdots & 0\\		
			\vdots &\vdots &\vdots &\vdots \\
			\eta_1 (\beta\alpha_1)^{k-1+t_1} & \eta_1 (\beta\alpha_2)^{k-1+t_1} & \cdots & \eta_1 (\beta\alpha_k)^{k-1+t_1} \\			
			\vdots &\vdots &\vdots &\vdots \\
			\eta_2 (\beta\alpha_1)^{k-1+t_2} & \eta_2 (\beta\alpha_2)^{k-1+t_2} & \cdots & \eta_2 (\beta\alpha_k)^{k-1+t_2} \\
			\vdots &\vdots &\vdots &\vdots \\
			\eta_{\ell} (\beta\alpha_1)^{k-1+t_{\ell}} & \eta_{\ell} (\beta\alpha_2)^{k-1+t_{\ell}} & \cdots & \eta_{\ell} (\beta\alpha_k)^{k-1+t_{\ell}} \\
			\vdots &\vdots &\vdots &\vdots \\
			0 & 0 & \cdots & 0\\
		\end{array}\right)
		\begin{array}{c}
			\textcolor{white}{0}\\
			\textcolor{white}{\vdots}\\		
			\leftarrow (h_1 +1)^{th} ~row \\			
			\textcolor{white}{\vdots} \\
			\leftarrow (h_2 +1)^{th} ~row \\
			\textcolor{white}{\vdots}\\
			\leftarrow (h_{\ell} +1)^{th}~ row \\
			\textcolor{white}{\vdots}\\
			\textcolor{white}{0}\\
		\end{array}$$
		we get $B_{\beta}B_{\beta}^T$ as the following matrix:
		\begin{equation}\label{bbt}
			\left(\begin{array}{ccccccc}
				0~~\cdots & 0 & \cdots& \cdots& \cdots & 0&~~~\cdots0\\	
				& & & \vdots & & & \\
				
				0~~\cdots & \eta_1^2 \underset{i=1}{\overset{k}{\sum}} (\beta\alpha_i)^{2k-2+t_1+t_1} & \cdots &\eta_1 \eta_2 \underset{i=1}{\overset{k}{\sum}} (\beta\alpha_i)^{2k-2+t_1+t_2} & \cdots & \eta_1\eta_{\ell} \underset{i=1}{\overset{k}{\sum}} (\beta\alpha_i)^{2k-2+t_1+t_{\ell}}&~~~\cdots0 \\			
				
				& & & \vdots & & & \\
				
				0~~\cdots & \eta_2 \eta_1 \underset{i=1}{\overset{k}{\sum}} (\beta\alpha_i)^{2k-2+t_2+t_1} & \cdots & \eta_2^2 \underset{i=1}{\overset{k}{\sum}} (\beta\alpha_i)^{2k-2+t_2+t_2} & \cdots & \eta_2\eta_{\ell} \underset{i=1}{\overset{k}{\sum}} (\beta\alpha_i)^{2k-2+t_2+t_{\ell}}&~~~\cdots0 \\	
				
				& & & \vdots & & & \\
				
				0~~\cdots & \eta_{\ell} \eta_1 \underset{i=1}{\overset{k}{\sum}} (\beta\alpha_i)^{2k-2+t_{\ell}+t_1} & \cdots & \eta_{\ell}\eta_2 \underset{i=1}{\overset{k}{\sum}} (\beta\alpha_i)^{2k-2+t_{\ell}+t_2} &\cdots & \eta_{\ell}^2 \underset{i=1}{\overset{k}{\sum}} (\beta\alpha_i)^{2k-2+t_{\ell}+t_{\ell}} &~~~\cdots0\\
				
				& & & \vdots & & & \\
				
				0~~\cdots & 0 & \cdots & \cdots &\cdots& 0&~~~\cdots0\\
			\end{array}\right)
		\end{equation}
		Notice that, by using (\ref{eq8}), the elements in $(h_i +1)^{th}$ row (for $1 \leq i \leq \ell$) of the above matrix are all zero except at most one element. Additionally, since $h_1>0$, the first row is all-zero. 
		Also, the above matrix is symmetric as well. 
		Next, we compute $A_{\beta}B_{\beta}^T + B_{\beta}A_{\beta}^T$ and get the matrix:
		\begin{table}[!hb]
			\tiny
			\begin{equation}\label{eq2}
				\left(\begin{array}{cccccccccc}
					0 & 0&\cdots & \underset{i=1}{\overset{k}{\sum}} \eta_1 \left(\beta\alpha_i\right)^{w_1} & \cdots & \underset{i=1}{\overset{k}{\sum}} \eta_2 \left(\beta\alpha_i\right)^{w_2} & \cdots &\underset{i=1}{\overset{k}{\sum}} \eta_{\ell} \left(\beta\alpha_i\right)^{w_{\ell}} &\cdots &0\\
					
					0 & 0& \cdots &\underset{i=1}{\overset{k}{\sum}} \eta_1 \left(\beta\alpha_i\right)^{w_1+1} & \cdots & \underset{i=1}{\overset{k}{\sum}} \eta_2 \left(\beta\alpha_i\right)^{w_2+1} & \cdots &\underset{i=1}{\overset{k}{\sum}} \eta_{\ell} \left(\beta\alpha_i\right)^{w_{\ell}+1} &\cdots &0\\

					\vdots&\vdots &\vdots &\vdots &\vdots &\vdots &\vdots &\vdots &\vdots &\vdots \\ 
					
					\underset{i=1}{\overset{k}{\sum}} \eta_1 \left(\beta\alpha_i\right)^{w_1} & \underset{i=1}{\overset{k}{\sum}} \eta_1 \left(\beta\alpha_i\right)^{w_1+1} & \cdots&0 & \cdots & \cdots & \cdots & \cdots & \cdots & \underset{i=1}{\overset{k}{\sum}} \eta_1 \left(\beta\alpha_i\right)^{w_1+k-1} \\
					
					\vdots &\vdots&\vdots &\vdots &\vdots &\vdots &\vdots &\vdots &\vdots &\vdots \\ 
					
					\underset{i=1}{\overset{k}{\sum}} \eta_2 \left(\beta\alpha_i\right)^{w_2} & \underset{i=1}{\overset{k}{\sum}} \eta_2 \left(\beta\alpha_i\right)^{w_2+1} & \cdots &\cdots & \cdots & 0 & \cdots & \cdots & \cdots & \underset{i=1}{\overset{k}{\sum}} \eta_2 \left(\beta\alpha_i\right)^{w_2+k-1} \\
					
					\vdots &\vdots&\vdots &\vdots &\vdots &\vdots &\vdots &\vdots &\vdots &\vdots \\ 
					
					\underset{i=1}{\overset{k}{\sum}} \eta_{\ell} \left(\beta\alpha_i\right)^{w_{\ell}} & \underset{i=1}{\overset{k}{\sum}} \eta_{\ell} \left(\beta\alpha_i\right)^{w_{\ell}+1} & \cdots&\cdots & \cdots & \cdots & \cdots & 0 & \cdots & \underset{i=1}{\overset{k}{\sum}} \eta_{\ell} \left(\beta\alpha_i\right)^{w_{\ell}+k-1} \\
					
					\vdots &\vdots&\vdots &\vdots &\vdots &\vdots &\vdots &\vdots &\vdots &\vdots \\ 
					
					0 &0&\vdots&\underset{i=1}{\overset{k}{\sum}} \eta_1 \left(\beta\alpha_i\right)^{w_1+k-1}&\cdots&\underset{i=1}{\overset{k}{\sum}} \eta_2 \left(\beta\alpha_i\right)^{w_2+k-1}&\cdots&\underset{i=1}{\overset{k}{\sum}} \eta_{\ell} \left(\beta\alpha_i\right)^{w_{\ell}+k-1}&\cdots&0\\
				\end{array}\right)
			\end{equation}
		\end{table}
	
		\noindent where $w_i= k-1+t_i $ for $1 \leq i \leq \ell$.
		Again, notice that, by using (\ref{eq8}), the elements in $(h_i +1)^{th}$ row (for $1 \leq i \leq \ell$) in the above matrix are all zero except at most one element. 
		Also, the above matrix is symmetric, with diagonal entries all-zero because $(h_i+1,h_i+1)^{th}$ entry is multiple of $2$, for each $i$.
		In addition, since $t_1>1$, the first row is all-zero. 
		Then, the generator matrix of multi-twisted RS code $\mathcal{C}_k(\bm\alpha,\bm{t},\bm{h},\bm{\eta})$,  given in (\ref{genmat}), can be written as $G =[A_1: A_{\gamma}] + [B_1: B_{\gamma}]$. Therefore $G \cdot G^T$ is given by 
		$$G \cdot G^T=\left([A_1: A_{\gamma}] + [B_1: B_{\gamma}]\right) \left( \left[ \begin{array}{c}
			A_1^T\\
			.~~ . \\
			A_{\gamma}^T
		\end{array}
		\right] + \left[ \begin{array}{c}
			B_1^T\\
			.~~ . \\
			B_{\gamma}^T
		\end{array}
		\right] \right).$$This implies that
		\begin{equation}\label{ggt}
			G \cdot G^T= \left(A_1A_1^T + A_{\gamma}A_{\gamma}^T\right) + \left(B_1B_1^T + B_{\gamma}B_{\gamma}^T\right) + \Big(A_{\gamma}B_{\gamma}^T + B_{\gamma}A_{\gamma}^T  + A_1B_1^T + B_1A_{1}^T \Big).
		\end{equation}
		By using $\beta=1+\gamma$ in the Equations: (\ref{aat}) and (\ref{bbt}), we get 
		\begin{equation}\label{eq3}
			A_1A_1^T + A_{\gamma}A_{\gamma}^T= 	\left(\begin{array}{ccccc}
				0 & 0 & \cdots & 0& 0\\
				0 & \cdots & \cdots &0& k \left(1+\gamma^k\right) \\
				0 & \cdots & \cdots & k \left(1+\gamma^k\right)&0 \\
				\vdots &\vdots &\dots &\vdots &\vdots \\
				0 & k \left(1+\gamma^k\right) & 0 & \cdots & 0 \\
			\end{array}\right)\end{equation}
		and $B_1B_1^T + B_{\gamma}B_{\gamma}^T=$
		\begin{equation}\label{eq4}\hspace{-3mm}
			\left(\begin{array}{ccccccc}
				0~~\cdots & 0 & \cdots& \cdots& \cdots & 0&~~~\cdots0\\	
				& & & \vdots & & & \\
				
				0~~\cdots & \eta_1^2 \underset{i=1}{\overset{k}{\sum}} ({(1+\gamma)}\alpha_i)^{2k-2+t_1+t_1} & \cdots &\eta_1 \eta_2 \underset{i=1}{\overset{k}{\sum}} ({(1+\gamma)}\alpha_i)^{2k-2+t_1+t_2} & \cdots & \eta_1\eta_{\ell} \underset{i=1}{\overset{k}{\sum}} ({(1+\gamma)}\alpha_i)^{2k-2+t_1+t_{\ell}}&~~~\cdots0 \\			
				
				& & & \vdots & & & \\
				
				0~~\cdots & \eta_2 \eta_1 \underset{i=1}{\overset{k}{\sum}} ({(1+\gamma)}\alpha_i)^{2k-2+t_2+t_1} & \cdots & \eta_2^2 \underset{i=1}{\overset{k}{\sum}} ({(1+\gamma)}\alpha_i)^{2k-2+t_2+t_2} & \cdots & \eta_2\eta_{\ell} \underset{i=1}{\overset{k}{\sum}} ({(1+\gamma)}\alpha_i)^{2k-2+t_2+t_{\ell}}&~~~\cdots0 \\	
				
				& & & \vdots & & & \\
				
				0~~\cdots & \eta_{\ell} \eta_1 \underset{i=1}{\overset{k}{\sum}} ({(1+\gamma)}\alpha_i)^{2k-2+t_{\ell}+t_1} & \cdots & \eta_{\ell}\eta_2 \underset{i=1}{\overset{k}{\sum}} ({(1+\gamma)}\alpha_i)^{2k-2+t_{\ell}+t_2} &\cdots & \eta_{\ell}^2 \underset{i=1}{\overset{k}{\sum}} ({(1+\gamma)}\alpha_i)^{2k-2+t_{\ell}+t_{\ell}} &~~~\cdots0\\
				
				& & & \vdots & & & \\
				
				0~~\cdots & 0 & \cdots & \cdots &\cdots& 0&~~~\cdots0\\
			\end{array}\right)
		\end{equation}
		\noindent Similarly, the remainder terms of (\ref{ggt}) can be obtained by 
		replacing $\beta$ with $(1+\gamma)$ in (\ref{eq2}).
		Then, by (\ref{ggt}) we deduce that
		$$G\cdot G^T=\left(\begin{array}{ccccc}
			0 & 0 & \cdots & 0& 0\\
			\\
			0 & \cdots & \cdots &0& k \left(1+\gamma^k\right) \\
			\\
			0 & \cdots & \cdots & k \left(1+\gamma^k\right)&0 \\
			\\
			\vdots &\vdots &\udots &\vdots &\vdots \\
			\\
			0 & k \left(1+\gamma^k\right) & 0 & \cdots & 0 \\
		\end{array}\right) +\left(\begin{array}{cccccccccccc}
			0& \cdots&  & 0&& &\cdots &0 \\
			\vdots&\textcolor{white}{ \left(\gamma^k\right) }& & \ast_1 & && &  \\
			
			&~~~~\ddots&&&\ast_2&&& \\
			
			0&\ast_1 & \Delta_1 &&& &&\\
			
			&&&\ddots&&&\ast_{\ell}& \\
			
			\vdots&\ast_2~~&&&\Delta_2&\\
			
			&&&&&\ddots&&\\
			
			& & & \ast_{\ell} & & &\Delta_{\ell} &\\
			
			0&&&&&&&\ddots \\
			
		\end{array}\right)_{k \times k}$$In the above equation, the right-most symmetric matrix has the first row all-zero, and 
		$\ast_i$ represents the possible non-zero entry for each $i$; 
		also, $\Delta_i$ is the possible non-zero entry at the $(h_i+1)^{th}$ diagonal position.
%		Since $k \vert (q-1)$, $k$ is odd, the diagonal entries $\Delta_1, \Delta_2,\dots, \Delta_{\ell}$ cannot coincide with the entries below the `reverse diagonal' consisting of $k(1+\gamma^k)$ from the first matrix.
%		Additionally $(1+ \gamma^k)\neq 0$.
		Clearly, $G\cdot G^T$ has rank at most $k-1$, and this completes the proof.
	\end{proof}
\end{theorem}

\begin{corollary}
	Let $q $ be a power of $2$, and $k>1$ be an integer such that $k \vert (q - 1)$. Then there exists an MDS $[2k,k]$ multi-twisted RS code $\mathcal{C}_k(\bm\alpha,\bm{t},\bm{h},\bm{\eta})$ over $\mathbb{F}_q$ with $\mathfrak{L}$-dimensional hull ($\mathfrak{L}>0$) for $\bm\alpha = (\alpha_1,\dots,\alpha_k,\gamma\alpha_1,\dots,\gamma\alpha_k)$, $h_1>0$ and $t_1>1$ if and only if $\bm \eta$ satisfies conditions of Theorem \ref{lemma4.1}.
\end{corollary}

\begin{example}
	Consider the finite field $ \mathbb{F}_{2^4}= \frac{\mathbb{F}_2[x]}{\left<x^4 + x +1\right>}$. Let $\gamma=\alpha$ be the primitive element as defined above, $k = 3$, $\bm t=(2, 3)$, $\bm h=(1, 2)$, and $\bm \eta = (\alpha^3, \alpha^3 + \alpha^2)$. Corresponding to these parameters, consider a $[6, 3, 4]$ double-twisted RS code $\mathcal{C}_{k,\bm t,\bm h,\bm \eta}$, where $\bm \alpha = (\alpha^2 + \alpha, \alpha^2 + \alpha + 1, 1, \alpha^3 + \alpha^2, \alpha^3 + \alpha^2 + \alpha, \alpha)$. Following to the proof of above theorem, we have 
	\begin{equation*}
		A_1A_1^T=\left(\begin{array}{ccc}
			1 & 0 & 0\\
			0 & 0 & 1\\
			0 & 1 & 0 \\
		\end{array}\right), ~~
		B_1 B_1^T=\left(\begin{array}{ccc}
			0 & 0 & 0\\
			0 & 0 & \alpha^3 + \alpha\\
			0 & \alpha^3 + \alpha & 0 \\
		\end{array}\right),
	\end{equation*}
	\begin{equation*}
		A_\alpha A_\alpha^T=\left(\begin{array}{ccc}
			1 & 0 & 0\\
			0 & 0 & \alpha^3\\
			0 & \alpha^3 & 0 \\
		\end{array}\right),~~
		B_\alpha B_\alpha^T=\left(\begin{array}{ccc}
			1 & 0 & 0\\
			0 & 0 & \alpha^3\\
			0 & \alpha^3 & 0 \\
		\end{array}\right).
	\end{equation*}
Therefore 
	\begin{equation*}
		A_1 A_1^T + A_\alpha A_\alpha^T =\left(\begin{array}{ccc}
			0 & 0 & 0\\
			0 & 0 & \alpha^3+1\\
			0 & \alpha^3+1 & 0 \\
		\end{array}\right),~~
		B_1 B_1^T + B_\alpha B_\alpha^T =\left(\begin{array}{ccc}
			0 & 0 & 0\\
			0 & 0 & \alpha\\
			0 & \alpha & 0 \\
		\end{array}\right),
	\end{equation*}
and
	\begin{equation*}
		A_\alpha B_\alpha^T + B_\alpha A_\alpha^T +A_1 B_1^T + B_1 A_1^T =
		\left(\begin{array}{ccc}
			0 & 0 & 0\\
			0 & 0 & 1\\
			0 & 1 & 0 \\
		\end{array}\right).
	\end{equation*}
Thus we can write
$$G\cdot G^T=\left(\begin{array}{ccc}
	0 & 0 & 0\\
	0 & 0& \alpha^3 + 1 \\
	0 & \alpha^3 + 1 & 0 \\
\end{array}\right) +
\left(\begin{array}{ccc}
	0 & 0 & 0\\
	0 & 0 & \alpha +1\\
	0 & \alpha +1 & 0 \\
\end{array}\right)$$which has rank $2$. Thus, $\mathcal{C}_{k,\bm t,\bm h,\bm \eta}$ is an MDS double-twisted RS code with one-dimensional hull.
\end{example}

\begin{theorem}\label{prop5.11}
	Let $q$ be a power of an odd prime, and $k>2$ be an integer such that $k \vert (q - 1)$.
	Then there exists a $[2k,k-1]$ multi-twisted RS code $\mathcal{C}_k(\bm\alpha,\bm{t},\bm{h},\bm{\eta})$ over $\mathbb{F}_q$ with $\mathfrak{L}$-dimensional hull ($\mathfrak{L}>0$) for $\bm\alpha = (\alpha_1,\dots,\alpha_k,\gamma\alpha_1,\dots,\gamma\alpha_k)$, $h_1>1$ and $t_\ell<k$.
	\begin{proof}
		Let, for $\beta\in \mathbb{F}_q^*$, $A_{\beta}$ be a $(k-1) \times k$ matrix over $\mathbb{F}_q$ given by $$A_{\beta}=\left(\begin{array}{cccc}
			1 & 1 & \cdots & 1\\
			\beta \alpha_1 & \beta \alpha_2 & \cdots & \beta \alpha_k \\
			(\beta \alpha_1)^2 & (\beta \alpha_2)^2 & \cdots & (\beta \alpha_k)^2 \\
			\vdots &\vdots &\vdots &\vdots \\
			(\beta \alpha_1)^{k-2} & (\beta \alpha_2)^{k-2} & \cdots & (\beta \alpha_k)^{k-2} \\
		\end{array}\right),$$then, by using (\ref{eq8}), we have a $(k-1) \times (k-1)$ matrix
		\begin{equation}\label{eq16}
			A_{\beta}A_{\beta}^T=\left(\begin{array}{ccccc}
				k & 0 & \cdots & 0&0\\
				0 & 0 & \cdots &0&0 \\
				0 & 0 & \cdots & 0&k \beta^k \\
				\vdots &\vdots &\dots &\udots&\vdots \\
				0 & 0 & k \beta^k& \cdots &0 \\
			\end{array}\right).
		\end{equation}
		Considering $B_{\beta}$ a $(k-1) \times k$ matrix over $\mathbb{F}_q$ as
		$$B_{\beta}=\left(\begin{array}{cccc}
			0 & 0 & \cdots & 0\\		
			\vdots &\vdots &\vdots &\vdots \\
			\eta_1 (\beta\alpha_1)^{k-1+t_1} & \eta_1 (\beta\alpha_2)^{k-1+t_1} & \cdots & \eta_1 (\beta\alpha_k)^{k-1+t_1} \\			
			\vdots &\vdots &\vdots &\vdots \\
			\eta_2 (\beta\alpha_1)^{k-1+t_2} & \eta_2 (\beta\alpha_2)^{k-1+t_2} & \cdots & \eta_2 (\beta\alpha_k)^{k-1+t_2} \\
			\vdots &\vdots &\vdots &\vdots \\
			\eta_{\ell} (\beta\alpha_1)^{k-1+t_{\ell}} & \eta_{\ell} (\beta\alpha_2)^{k-1+t_{\ell}} & \cdots & \eta_{\ell} (\beta\alpha_k)^{k-1+t_{\ell}} \\
			\vdots &\vdots &\vdots &\vdots \\
			0 & 0 & \cdots & 0\\
		\end{array}\right)
		\begin{array}{c}
			\textcolor{white}{0}\\
			\textcolor{white}{\vdots}\\		
			\leftarrow (h_1 +1)^{th} ~row \\			
			\textcolor{white}{\vdots} \\
			\leftarrow (h_2 +1)^{th} ~row \\
			\textcolor{white}{\vdots}\\
			\leftarrow (h_{\ell} +1)^{th}~ row \\
			\textcolor{white}{\vdots}\\
			\textcolor{white}{0}\\
		\end{array}$$
		we get $B_{\beta}B_{\beta}^T$ as the following matrix:
		\begin{equation}\label{bbtt}
			\left(\begin{array}{ccccccc}
				0~~\cdots & 0 & \cdots& \cdots& \cdots & 0&~~~\cdots0\\	
				& & & \vdots & & & \\
				
				0~~\cdots & \eta_1^2 \underset{i=1}{\overset{k}{\sum}} (\beta\alpha_i)^{2k-2+t_1+t_1} & \cdots &\eta_1 \eta_2 \underset{i=1}{\overset{k}{\sum}} (\beta\alpha_i)^{2k-2+t_1+t_2} & \cdots & \eta_1\eta_{\ell} \underset{i=1}{\overset{k}{\sum}} (\beta\alpha_i)^{2k-2+t_1+t_{\ell}}&~~~\cdots0 \\			
				
				& & & \vdots & & & \\
				
				0~~\cdots & \eta_2 \eta_1 \underset{i=1}{\overset{k}{\sum}} (\beta\alpha_i)^{2k-2+t_2+t_1} & \cdots & \eta_2^2 \underset{i=1}{\overset{k}{\sum}} (\beta\alpha_i)^{2k-2+t_2+t_2} & \cdots & \eta_2\eta_{\ell} \underset{i=1}{\overset{k}{\sum}} (\beta\alpha_i)^{2k-2+t_2+t_{\ell}}&~~~\cdots0 \\	
				
				& & & \vdots & & & \\
				
				0~~\cdots & \eta_{\ell} \eta_1 \underset{i=1}{\overset{k}{\sum}} (\beta\alpha_i)^{2k-2+t_{\ell}+t_1} & \cdots & \eta_{\ell}\eta_2 \underset{i=1}{\overset{k}{\sum}} (\beta\alpha_i)^{2k-2+t_{\ell}+t_2} &\cdots & \eta_{\ell}^2 \underset{i=1}{\overset{k}{\sum}} (\beta\alpha_i)^{2k-2+t_{\ell}+t_{\ell}} &~~~\cdots0\\
				
				& & & \vdots & & & \\
				
				0~~\cdots & 0 & \cdots & \cdots &\cdots& 0&~~~\cdots0\\
			\end{array}\right)
		\end{equation}
		Notice that, by using (\ref{eq8}), the elements in $(h_i +1)^{th}$ row (for $1 \leq i \leq \ell$) of the above matrix are all zero except at most one element. Additionally, since $h_1>1$, the first and second rows are all-zero. 
		Also, the above matrix is symmetric as well.
		Next, we compute $A_{\beta}B_{\beta}^T + B_{\beta}A_{\beta}^T$ and get the matrix:
		\begin{table}[!hb]
			\tiny
			\begin{equation}\label{eq18}
				\left(\begin{array}{cccccccccc}
					0 & 0&\cdots & \underset{i=1}{\overset{k}{\sum}} \eta_1 \left(\beta\alpha_i\right)^{w_1} & \cdots & \underset{i=1}{\overset{k}{\sum}} \eta_2 \left(\beta\alpha_i\right)^{w_2} & \cdots &\underset{i=1}{\overset{k}{\sum}} \eta_{\ell} \left(\beta\alpha_i\right)^{w_{\ell}} &\cdots &0\\
					
					0 & 0& \cdots &\underset{i=1}{\overset{k}{\sum}} \eta_1 \left(\beta\alpha_i\right)^{w_1+1} & \cdots & \underset{i=1}{\overset{k}{\sum}} \eta_2 \left(\beta\alpha_i\right)^{w_2+1} & \cdots &\underset{i=1}{\overset{k}{\sum}} \eta_{\ell} \left(\beta\alpha_i\right)^{w_{\ell}+1} &\cdots &0\\

					\vdots&\vdots &\vdots &\vdots &\vdots &\vdots &\vdots &\vdots &\vdots &\vdots \\ 
					
					\underset{i=1}{\overset{k}{\sum}} \eta_1 \left(\beta\alpha_i\right)^{w_1} & \underset{i=1}{\overset{k}{\sum}} \eta_1 \left(\beta\alpha_i\right)^{w_1+1} & \cdots&2\underset{i=1}{\overset{k}{\sum}} \eta_1 \left(\beta\alpha_i\right)^{w_1 +h_1} & \cdots & \cdots & \cdots & \cdots & \cdots & \underset{i=1}{\overset{k}{\sum}} \eta_1 \left(\beta\alpha_i\right)^{w_1+k-1} \\
					
					\vdots &\vdots&\vdots &\vdots &\vdots &\vdots &\vdots &\vdots &\vdots &\vdots \\ 
					
					\underset{i=1}{\overset{k}{\sum}} \eta_2 \left(\beta\alpha_i\right)^{w_2} & \underset{i=1}{\overset{k}{\sum}} \eta_2 \left(\beta\alpha_i\right)^{w_2+1} & \cdots &\cdots & \cdots & 2\underset{i=1}{\overset{k}{\sum}} \eta_2 \left(\beta\alpha_i\right)^{w_1 +h_2} & \cdots & \cdots & \cdots & \underset{i=1}{\overset{k}{\sum}} \eta_2 \left(\beta\alpha_i\right)^{w_2+k-1} \\
					
					\vdots &\vdots&\vdots &\vdots &\vdots &\vdots &\vdots &\vdots &\vdots &\vdots \\ 
					
					\underset{i=1}{\overset{k}{\sum}} \eta_{\ell} \left(\beta\alpha_i\right)^{w_{\ell}} & \underset{i=1}{\overset{k}{\sum}} \eta_{\ell} \left(\beta\alpha_i\right)^{w_{\ell}+1} & \cdots&\cdots & \cdots & \cdots & \cdots & 2\underset{i=1}{\overset{k}{\sum}} \eta_\ell \left(\beta\alpha_i\right)^{w_1 +h_\ell} & \cdots & \underset{i=1}{\overset{k}{\sum}} \eta_{\ell} \left(\beta\alpha_i\right)^{w_{\ell}+k-1} \\
					
					\vdots &\vdots&\vdots &\vdots &\vdots &\vdots &\vdots &\vdots &\vdots &\vdots \\ 
					
					0 &0&\vdots&\underset{i=1}{\overset{k}{\sum}} \eta_1 \left(\beta\alpha_i\right)^{w_1+k-1}&\cdots&\underset{i=1}{\overset{k}{\sum}} \eta_2 \left(\beta\alpha_i\right)^{w_2+k-1}&\cdots&\underset{i=1}{\overset{k}{\sum}} \eta_{\ell} \left(\beta\alpha_i\right)^{w_{\ell}+k-1}&\cdots&0\\
				\end{array}\right)
			\end{equation}
		\end{table}
		
		\noindent where $w_i=k-1+t_i$ for $1 \leq i \leq \ell$. 
		Again, notice that, by using (\ref{eq8}), the elements in $(h_i +1)^{th}$ row (for $1 \leq i \leq \ell$) in the above matrix are all zero except at most one element. 
		Also, the above matrix is symmetric. 
		In addition, since $t_\ell<k$, the second row is all-zero. 
		Then, the generator matrix of multi-twisted RS code $\mathcal{C}_k(\bm\alpha,\bm{t},\bm{h},\bm{\eta})$, given in (\ref{genmat}), can be written as $G =[A_1: A_{\gamma}] + [B_1: B_{\gamma}]$. 
		Similar to the computations as in Theorem \ref{prop4.2}, the $(k-1)\times (k-1)$ matrix $G \cdot G^T$ is given by the following expression:
		$$\left(\begin{array}{cccccc}
			2k &0& 0 & \cdots & 0& 0\\
			
			0 &0& 0 & \cdots & 0& 0\\
			\\
			0 &0& \cdots & \cdots &0& k \left(1+\gamma^k\right) \\
			\\
			0 & 0&\cdots & \cdots & k \left(1+\gamma^k\right)&0 \\
			\vdots &\vdots &\vdots &\udots &\vdots \\
			\\
			0 &0& k \left(1+\gamma^k\right) & 0 & \cdots & 0 \\
		\end{array}\right) +\left(\begin{array}{cccccccccc}
			
			{ \textcolor{white}{.}}& 0&  & \ast_1&& &  \\
			0&0&\cdots&&&&\cdots&0\\
			&\vdots&\ddots&&&\ast_2&& \\
			
			\ast_1 &&& \Delta_1 &&& \textcolor{white}{\left(\gamma^k\right)}\\
			
			&&&&\ddots&&&\ast_{\ell} \\
			
			&0&\ast_2&&&\Delta_2&\\
			
			&&&&&~~\ddots&&\\
			
			&\vdots & & \ast_{\ell} & & &\Delta_{\ell}\\
			
			&0&&&&&&\ddots \\
			
		\end{array}\right)$$In the above expression, the right-most symmetric matrix has the second row all-zero, and  $\ast_i$ represents the  possible non-zero entry for each $i$; also, since $h_1>1$, $\Delta_1$ cannot occur in first and second rows.
		Clearly, $G\cdot G^T$ has rank at most $k-2$, and this completes the proof.
	\end{proof}
\end{theorem}

\begin{corollary}
	Let $q$ be a power of an odd prime, and $k>2$ be an integer such that $k \vert (q - 1)$. 
	Then there exists an MDS $[2k,k-1]$ multi-twisted RS code $\mathcal{C}_k(\bm\alpha,\bm{t},\bm{h},\bm{\eta})$ over $\mathbb{F}_q$ with $\mathfrak{L}$-dimensional hull ($\mathfrak{L}>0$) for $\bm\alpha = (\alpha_1,\dots,\alpha_k,\gamma\alpha_1,\dots,\gamma\alpha_k)$, $h_1>1$, $t_\ell<k$ if and only if $\bm \eta$ satisfies the conditions of Theorem \ref{lemma4.1}.
\end{corollary}

\begin{example}
	Consider the finite field $ \mathbb{F}_{3^4}= \frac{\mathbb{F}_2[x]}{\left<x^4 + 2x^3 +2\right>}$. Let $\gamma=\alpha$ be the primitive element as defined above, $k = 5$, $\bm t=(1, 2)$, $\bm h=(2, 3)$, and $\bm \eta = (\alpha^3+\alpha^2, \alpha)$. 
	Corresponding to these parameters, consider the $[10, 4, 7]$ double-twisted RS code $\mathcal{C}_{k,\bm t,\bm h,\bm \eta}$, where $\bm \alpha = (2\alpha^2 + \alpha+2, 2 \alpha^3 + \alpha + 2, 2 \alpha^2 + 2 \alpha + 1, \alpha^3 + 2 \alpha^2 + 2 \alpha, 1, 2 \alpha^3 + \alpha^2 + 2 \alpha, 2 \alpha^3 + \alpha^2 + 2 \alpha + 2, 2 \alpha^3 + 2 \alpha^2 + \alpha, 2 \alpha^2 + 1, \alpha)$. Following to the proof of above theorem, we have 
	\begin{equation*}
		A_1A_1^T=\left(\begin{array}{cccc}
			2 & 0 & 0 & 0\\
			0 & 0 & 0 & 0\\
			0 & 0 & 0 & 2\\
			0 & 0 & 2 & 0\\
		\end{array}\right),~~
		B_1 B_1^T=\left(\begin{array}{cccc}
			0 & 0 & 0 & 0\\
			0 & 0 & 0 & 0\\
			0 & 0 & 2\alpha^3 + 2\alpha^2 + 2 & 0\\
			0 & 0 & 0 & 0\\
		\end{array}\right),
	\end{equation*}
	\begin{equation*}
		A_\alpha A_\alpha^T=\left(\begin{array}{cccc}
			2 & 0 & 0 & 0\\
			0 & 0 & 0 & 0\\
			0 & 0 & 0 & 2 \alpha^3 + 2 \alpha + 2\\
			0 & 0 & 2 \alpha^3 + 2 \alpha + 2 & 0\\
		\end{array}\right),~~
		B_\alpha B_\alpha^T=\left(\begin{array}{cccc}
			0 & 0 & 0 & 0\\
			0 & 0 & 0 & 0\\
			0 & 0 & \alpha^3 +  \alpha^2 & 0\\
			0 & 0 & 0 & 0\\
		\end{array}\right).
	\end{equation*}
	Therefore
	\begin{equation*}
		A_1 A_1^T + A_\alpha A_\alpha^T =\left(\begin{array}{cccc}
			1 & 0 & 0 & 0\\
			0 & 0 & 0 & 0\\
			0 & 0 & 0 & 2 \alpha^3 + 2 \alpha + 1\\
			0 & 0 & 2 \alpha^3 + 2 \alpha + 1 & 0 \\
		\end{array}\right), ~~
		B_1 B_1^T + B_\alpha B_\alpha^T =\left(\begin{array}{cccc}
			0 & 0 & 0 & 0\\
			0 & 0 & 0 & 0\\
			0 & 0 & 2 & 0\\
			0 & 0 & 0 & 0\\
		\end{array}\right),
	\end{equation*}
	and
	\begin{equation*}
		A_\alpha B_\alpha^T + B_\alpha A_\alpha^T +A_1 B_1^T + B_1 A_1^T =
		\left(\begin{array}{cccc}
			0 & 0 &  \alpha & 0\\
			0 & 0 & 0 & 0\\
			 \alpha & 0 & 0 & 0\\
			0 & 0 & 0 & 0\\
		\end{array}\right).
	\end{equation*}
	Thus we can write
	$$G\cdot G^T=\left(\begin{array}{cccc}
		1 & 0 & 0 & 0\\
		0 & 0 & 0 & 0\\
		0 & 0 & 0 & 2 \alpha^3 + 2 \alpha + 1\\
		0 & 0 & 2 \alpha^3 + 2 \alpha + 1 & 0 \\
	\end{array}\right) +
	\left(\begin{array}{cccc}
		0 & 0 &  \alpha& 0\\
		0 & 0 & 0 & 0\\
		 \alpha & 0 & 2 & 0\\
		0 & 0 & 0 & 0\\
	\end{array}\right)$$
	which have rank $3$. Thus, $\mathcal{C}_{k,\bm t,\bm h,\bm \eta}$ is an MDS double-twisted RS code with one-dimensional hull.
\end{example}

\section{Conclusion}\label{sec6}
The construction of MDS codes is one of the active and hot research topics in the area of algebraic coding theory due to their maximum error-correction property. We obtained necessary and sufficient condition for multi-twisted RS codes to be MDS. Particularly, we focused MDS double-twisted RS codes. In future, one can study the existence of the error-correcting pair \cite{he_2023} for these double-twisted RS codes. We also studied multi-twisted RS codes with small dimensional hull. Further, we obtained necessary and sufficient conditions for such MDS multi-twisted RS codes to have small dimensional hull. As a future task, one can study such (MDS) multi-twisted RS codes with certain dimensional Hermitian hull \cite{Fang_2020} and Galois hull \cite{Cao_2021, Fang_2022, Liu2020} with some possible applications.
For comparison of these studies with existing studies, we refer to the Table \ref{table::comparison}.
\begin{table}[t!h]
	\centering
	\begin{tabular}{|p{6cm}|p{6.5cm}|p{2cm}|}
		\hline
		RS codes (always MDS) &\centering $\eta=0$ & \cite{Reed1960}\\ \hline
		Single-twisted RS codes &\centering $\eta \neq 0$, $1\leq t \leq n-k$, $0\leq h\leq k-1$& \cite{Beelen2017}\\
		\hline
		Multi-twisted RS codes & \centering$0\leq h_i<  h_{i+1}\leq k-1$,
		$1\leq t_i<t_{i+1} \leq n - k$,
		$\bm t = (t_1,t_2,\dots,t_\ell)$,
		$\bm h=(h_1,h_2,\dots,h_\ell)$,
		$\bm\eta = (\eta_1,\eta_2,\dots,\eta_\ell )\in (\mathbb{F}_q^*)^\ell$ & \cite{Beelen2018a}\\ \hline
		Necessary and sufficient condition for single-twisted RS codes to be MDS &\centering$\eta\neq0$, $1 \leq t \leq n-k$, $0 \leq h \leq k-1$&\cite{Beelen2017, Sui_2022}\\ \hline
		Necessary and sufficient condition for double-twisted RS codes to be MDS&\centering $\bm\eta=(\eta_1,\eta_2)\in (\mathbb{F}_q^*)^2$, $\bm t = (1,2)$, $\bm h = (k-1,k-2)$&\cite{Sui_2022a}\\ \hline
		Necessary and sufficient condition for double-twisted RS codes to be MDS&\centering$\bm\eta=(\eta_1,\eta_2)\in (\mathbb{F}_q^*)^2$, $\bm t = (1,2)$, $\bm h = (0,1)$&In this paper\\ \hline
		Necessary and sufficient condition for Multi-twisted RS codes to be MDS&\centering $\bm t = (1,2,\dots, \ell)$, $\bm h = (k-\ell, k-\ell+1, \dots, k-1)$, $\bm \eta =(\eta_1,\eta_2,\dots,\eta_\ell)\in (\mathbb{F}_q^*)^\ell$, $\ell< \min\{k,n-k\}$&\cite{Gu}\\ \hline
		Necessary and sufficient condition for Multi-twisted RS codes to be MDS&\centering$0\leq h_i<  h_{i+1}\leq k-1$,
		$1\leq t_i<t_{i+1} \leq n - k$,
		$\bm t = (t_1,t_2,\dots,t_\ell)$,
		$\bm h=(h_1,h_2,\dots,h_\ell)$,
		$\bm\eta = (\eta_1,\eta_2,\dots,\eta_\ell )\in (\mathbb{F}_q^*)^\ell$&In this paper\\
		\hline
		\hline
		Explicit construction for MDS single-twisted RS codes&\centering $\eta\neq0$, $t=1$, $h=0,k-1$&\cite{Beelen2017}\\
		\hline
		Explicit construction for MDS single-twisted RS codes&\centering $\eta\neq0$, $t=2$, $h=1$&\cite{Sui_2022}\\
		\hline
		Explicit construction for MDS double-twisted RS codes&\centering $\bm\eta=(\eta_1,\eta_2)\in (\mathbb{F}_q^*)^2$, $\bm t = (1,2)$, $\bm h = (0,1)$&In this paper\\
		\hline \hline
		Single-twisted RS codes with one-dimensional hull&\centering $\eta \neq 0$; $1\leq t \leq n-k$; $0\leq h\leq k-1$& \cite{Wu2021} \\
		\hline
		MDS Single-twisted RS codes with zero-dimensional hull&\centering $\eta\neq0$, $t=1$, $h=k-1$&\cite{Liu2021}\\
		\hline
		Multi-twisted RS codes with zero-dimensional hull&\centering $\bm t = (1,2,\dots, \ell)$, $\bm h = (0,1,\dots,\ell-1)$, $\bm \eta =(\eta_1,\eta_2,\dots,\eta_\ell)\in (\mathbb{F}_q^*)^\ell$, $\ell\leq n/2$&\cite{Liu2021}\\ \hline
		Multi-twisted RS codes with small dimensional hull & \centering$0\leq h_i<  h_{i+1}\leq k-1$,
		$1\leq t_i<t_{i+1} \leq n - k$,
		$\bm t = (t_1,t_2,\dots,t_\ell)$,
		$\bm h=(h_1,h_2,\dots,h_\ell)$,
		$\bm\eta = (\eta_1,\eta_2,\dots,\eta_\ell )\in (\mathbb{F}_q^*)^\ell$ & In this paper\\ 
		\hline
		Necessary and sufficient conditions for MDS multi-twisted RS codes with small dimensional hull&\centering$1 < h_i <  h_{i+1}\leq k-1$,
		$1 < t_i<t_{i+1} \leq n - k-2$,
		$\bm t = (t_1,t_2,\dots,t_\ell)$,
		$\bm h=(h_1,h_2,\dots,h_\ell)$,
		$\bm\eta = (\eta_1,\eta_2,\dots,\eta_\ell )\in (\mathbb{F}_q^*)^\ell$ & In this paper\\ 
		\hline
	\end{tabular}\caption{\small Comparison with existing studies}\label{table::comparison}
\end{table}

\section*{Statements and Declarations}
\begin{itemize}
	\item \textbf{Acknowledgment:}
	We are extremely thankful and appreciate the referee(s) for his/her comments which led improvements towards the quality of the paper. In addition, we are grateful to Dr. S.~K. Tiwari for his fruitful discussions and suggestions towards improvement of the paper. 
	\item \textbf{Ethical approval:}
	The submitted work is original and not submitted to more than one journal for simultaneous consideration.
	\item \textbf{Competing interest:} None of the authors have any relevant financial or non-financial competing interests.
	\item \textbf{Author's contributions:} The conceptualization, methodology, investigation, writing-original draft preparation, review and revision-editing have been performed by both the authors equally.	
	
	\item \textbf{Funding:} This work has no financial supported.
	\item \textbf{Availability of data and materials:}
	This manuscript has no associated data.

\end{itemize}

\noindent Scientific Analysis Group\\
Metcalfe House Complex\\
Defence Research and Development Organisation\\
Delhi - 110054, India.\\
E-mails: harshdeep.sag@gov.in \{H. Singh\}\\
meenakapishchand@gmail.com \{K. C. Meena\}
%\newpage
\begin{appendices}
	\section{Implementation of Example \ref{ex::3.2}}\label{Appendix1}
	\noindent For checking the non-singularity of each matrix in Example \ref{ex::3.2}, we attach the following implementation:
	\begin{lstlisting}
F.<a> = GF(16)
alpha_vec = [0, a^2, a + 1, a^2 + a, a^3 + a + 1] 
eta1 = a^3 + a^2
eta2_list = [1, a^2 + 1, a^2 + a + 1, a^3, a^3 + a^2, a^3 + a^2 + a] 
R.<x> = PolynomialRing(F)
for eta2 in eta2_list:
	print('\n eta2 : ',eta2)
	det_matrix = []
	for I in Combinations([0,1,2,3,4],3):
		temp_poly = R(1)
		for i in I:
			temp_poly = temp_poly*(x-alpha_vec[i])
	sigma_coeff = temp_poly.coefficients(sparse= False)
	A_I = matrix(F,[[1,0],[sigma_coeff[2],1]])
	B_I = matrix(F,[[-1*sigma_coeff[0], -1*sigma_coeff[1]],[0,-1*sigma_coeff[0]]])
	D = diagonal_matrix(F,[eta2^(-1), eta1^(-1)])
	det_matrix.append((D*A_I + B_I).determinant())
	print(det_matrix)
	\end{lstlisting}
	The output of this code is described below:
	\begin{lstlisting}
eta2 : 1
[a^3 + a^2 + 1, a^3 + a^2 + a, a^3 + a + 1, a^2, a^3 + a^2 + a + 1, a^3 + 1, a, a^3 + a, a^3, a^2]

eta2 : a^2 + 1
[a^2 + 1, a^2 + a, a + 1, a^3 + a^2, a^2 + a + 1, 1, a^3 + a^2, a^3 + 1, a^2 + 1, a^3 + a + 1]

eta2 : a^2 + a + 1
[a^3 + a^2 + a, a^3 + a^2 + 1, a^3, a^2 + a + 1, a^3 + a^2, a^3 + a, a^3 + a^2 + 1, a^3 + a^2, 1, a^3 + 1]

eta2 : a^3
[a^3 + a + 1, a^3, a^3 + a^2 + 1, a, a^3 + 1, a^3 + a^2 + a + 1, a^3 + a^2 + a + 1, a^2 + a, a^3 + 1, a^3 + a^2 + 1]

eta2 : a^3 + a^2
[a^3 + a^2 + a + 1, a^3 + a^2, a^3 + 1, a^2 + a, a^3 + a^2 + 1, a^3 + a + 1, a^3, a^3 + a^2 + a, a^2 + a, a + 1]

eta2 : a^3 + a^2 + a
[a^3 + a, a^3 + 1, a^3 + a^2, a + 1, a^3, a^3 + a^2 + a, a^3 + a, a^2, a^3 + a^2 + a, a^2 + a + 1]
	\end{lstlisting}
\end{appendices}
\pagebreak
\begin{appendices}
\section{Implementation for counting of double-twisted RS codes}\label{Appendix2}
	The following SageMath implementation provides number of double-twisted RS codes of length $n$ and dimension $k$ over the finite field $\mathbb{F}_{q}$.

\begin{lstlisting}
def number(q,n,k):
	F.<a> = GF(q)
	A = Combinations(list(F),n)
	main_count = 0
	for C in A:
		count = 0
		for eta in F^2:
			if eta[0]*eta[1] !=0:
				Ik = Combinations(C, k)
				i = 0
				while i < len(Ik):
					sum_prod = 0
					for Ik1 in Combinations(Ik[i], k-1):
						sum_prod = sum_prod + prod(Ik1)
					flag = 1 - eta[0]*((-1)^k)*prod(Ik[i]) + eta[1]*((-1)^k)*((sum_prod)*(sum(Ik[i])) - prod(Ik[i])) + eta[0]*eta[1]*(prod(Ik[i])^2)
					i = i+1
					if flag==0:
						break
				if flag!=0:
					count = count + 1
		main_count = main_count + count
	return main_count                                                                                                                                                                                                                                                                                                                                                                                                                                                                                                                                                                                                                                                                                                                                                                                                                                                                                                                                                                                                                                                                                                                                                                                                                                                                                                                                                                                                                                 
\end{lstlisting}
\end{appendices}
\end{document}